\documentclass[twocol]{ametsocV6.1}

\usepackage{booktabs}
\usepackage{longtable}
\usepackage{xspace}
\usepackage{physics}
\usepackage{etoolbox}
\usepackage{array}
\usepackage{needspace}
\usepackage{tabularx}
\usepackage{caption}
\usepackage[T1]{fontenc}

\newcommand{\ie}{\textit{i}.\textit{e}.,\@\xspace}
\newcommand{\eg}{\textit{e}.\textit{g}.,\@\xspace}

\makeatletter
\newcommand\footnoteref[1]{\protected@xdef\@thefnmark{\ref{#1}}\@footnotemark}
\makeatother

\title{A Radar-Based Hail Climatology of Australia}

\authors{Jordan P. Brook, \aff{a,b}\correspondingauthor{Jordan Brook, j.brook@uq.edu.au}
Joshua S. Soderholm,\aff{b} 
Alain Protat,\aff{b} 
Hamish McGowan,\aff{a} 
Robert A. Warren\aff{b} 
}

\affiliation{\aff{a}{Atmospheric Observations Research Group, University of Queensland, QLD, Australia}\\
\aff{b}{Science and Innovation Group, Australian Bureau of Meteorology, Docklands, VIC, Australia}}

\abstract{In Australia, hailstorms present considerable public safety and economic risks, where they are considered the most damaging natural hazard in terms of annual insured losses. Despite these impacts, the current climatological distribution of hailfall across the continent is still comparatively poorly understood. This study aims to supplement previous national hail climatologies, such as those based on environmental proxies or satellite radiometer data, with more direct radar-based hail observations. The heterogeneous and incomplete nature of the Australian radar network complicates this task and prompts the introduction of some novel methodological elements. We introduce an empirical correction technique to account for hail reflectivity biases at C-band, derived by comparing overlapping C- and S-band observations. Furthermore, we demonstrate how object-based hail swath analysis may be used to produce resolution-invariant hail frequencies, and describe an interpolation method used to create a spatially continuous hail climatology. The Maximum Estimated Size of Hail (MESH) parameter is then applied to a mixture of over fifty operational radars in the Australian radar archive, resulting in the first nationwide, radar-based hail climatology. The spatiotemporal distribution of hailstorms is examined, including their physical characteristics, seasonal and diurnal frequency, and regional variations of such properties across the continent.} 

\begin{document}

\maketitle

\section{Introduction}

Hailstorms account for a substantial portion of financial impacts caused by natural hazards globally, resulting in greater than 10 billion USD in insured losses annually in the United States alone \citep{Gunturi17}. In Australia, hail is typically considered the most costly natural hazard in terms of insured losses, exceeding the combined losses of both bushfires and floods \citep{Schuster06, Mcaneney19}. Several of the country's most populous cities have experienced hailstorms that resulted in over 1 billion AUD in losses, including Melbourne in the south \citep{Buckley10, Allen12}, Perth on the west coast \citep{Buckley10}, and most frequently Brisbane and Sydney on the east coast \citep{Soderholm19, Schuster06}. Financial losses also appear to be increasing alongside an expansion of the built environment \citep{Changnon09}, and as of 2019 in Australia, seven of the ten most damaging events occurred in the past decade \citep[adjusted for inflation,][]{Munich19}. The significance of these financial impacts underscores the need to quantify the current level of hail hazard in the region.

Numerous studies have investigated the climatology of hailstorms in Australia using various data sources, including hail reports, hail proxies from weather models, and remote sensing observations from satellite radiometers and weather radars. First, studies based on the Bureau of Meteorology's Severe Storms Archive are limited by well-documented reporting biases for hail observations \citep{Allen15}. Population biases are particularly prominent in Australia, which has the fifth lowest population density globally. These problems are compounded by the fact that 87\% of the already limited population resides within 50 km of the coast, leaving vast areas (mostly in the country's arid interior) effectively unobserved \citep{ABS21}. Consequently, climatological studies based on hail reports are restricted to more populated regions such as eastern New South Wales \citep{Schuster05}. Changes in reporting practices further diminish the practical value of this data set \citep{Walsh16}, especially for more recent years where the archive has been poorly maintained. The most comprehensive analysis of hail reports in Australia is provided by \citet{Allen16}, offering valuable verification for the results that follow.

Perhaps the most common approach for estimating the climatological distribution of hail on regional scales involves identifying ``hail-prone'' environmental conditions from numerical weather prediction (NWP) models. In these studies, an environment is considered ``hail-prone'' based on a combination of favorable model indices, such as Convective Available Potential Energy (CAPE) or 0-6 km bulk vertical wind shear \citep[see][and references therein]{Raupach23}. Such an approach is appealing from a climatological standpoint due to the spatiotemporal continuity of the underlying model data, and has formed the basis of several global \citep[\eg][]{Prein18} and Australian \citep[\eg][]{Allen14} hail and severe storm climatologies. While these studies have contributed valuable insights into the distribution of hail events worldwide, they must be interpreted with some caution. First, the exact formation of the ``optimal'' hail proxy varies considerably between studies and different climatic regions \citep{Zhou21}, often resulting in an overestimation of the tropical hail frequency in Australia \citep{Bednarczyk12, Allen14, Prein18}. Recent research by \citet{Raupach23} addresses these concerns in an Australian context by training their proxy on Australian hail reports, and allowing their formulation to vary latitudinally, which more accurately characterized hail environments across the country. However, the accuracy of this approach is limited by the representativeness of the underlying hail report data set, which is affected by the aforementioned biases involving population density and reporting practices. Moreover, hail proxies are known to suffer from the ``initiation problem'', whereby only a small fraction of hail-prone environments \citep[often less than 10\%,][]{ Taszarek20} actually undergo convective initiation \citep{Tippett15}. This issue ultimately leads to large uncertainties regarding the true climatological frequency of hailfall. In an Australian context, this problem spuriously highlights a spatial hail frequency maximum over the ocean on the east coast \citep[\eg][]{Prein18}. These oceanic overestimations likely stem from an absence of known convection trigger mechanisms for the region, such as sea-breeze or topographic interactions \citep{Soderholm17b}. \citet{Raupach23} demonstrated that incorporating a series of additional environmental constraints could successfully reduce the false alarm rate for Australian hailstorms, improving the Critical Success Index (CSI) to $\sim$0.3 when compared to hail reports. 

Recent climatological studies also attempt to address the ``initiation problem'' by confirming convective initiation through passive spaceborne observations of deep convection. While passive infrared imagers aboard geostationary satellites cannot detect hail directly \citep{Murillo19}, some studies have used these observations to identify dynamical characteristics associated with hailstorms, such as above anvil cirrus plumes \citep{Levizzani96, Homeyer17} or overshooting tops \citep[OT,][]{Schmetz97, Bedka10}. OT detections have been combined with hail proxies to produce regional hail climatologies in numerous studies \citep[\eg][]{Bedka18b, Punge21}, and although this methodology likely improves on those derived from hail proxies alone, significant uncertainty still remains when associating OT detections with hail at the surface. \citet{Bedka18b} found that only 28\% of OT detections are spatiotemporally associated with hail reports in Australia (within 1 hour and 50 km), which is also considerably lower than ~50\% in Europe \citep{Bedka11} and the United States \citep{Dworak12}. Despite these uncertainties, the \citet{Bedka18b} hail climatology of Australia is a significant advancement in our understanding of hail in remote areas of the continent, and serves as a valuable reference for this study.

One final class of satellite-based hail climatologies relevant for our investigation involves hail detections from passive microwave radiometers and precipitation (Ku/Ka band) radars aboard low-earth orbit satellites. Studies have shown that simple brightness temperature (BT) thresholds at certain frequencies correctly subset hail-bearing storms in the US with ~40\% accuracy \citep{Cecil09, Ferraro15}. This principle has been employed to generate several global hail climatologies \citep[\eg][]{Cecil09, Cecil12, Ferraro15}. Subsequent research with spaceborne precipitation radars by \cite{Ni17} revealed that similar to hail proxies, these passive microwave methods overestimate the amount of hail in tropical regions such as northern Australia due to regional variations in thunderstorm structure. These insights have been used to more accurately detect hail events \citep[$\text{CSI}\approx 0.28$ relative to ground-radar-based hail detections,][]{Mroz17}, and updated global hail climatologies based on these improvements will also serve as important secondary validation for our study \citep{Mroz17, Ni17, Bang19}. 

Indirect hail detection methods from hail proxies, geostationary satellites, or passive microwave radiometers each have distinct limitations, which ultimately contributes to uncertainty in their resulting climatologies. In light of these constraints, weather radars offer an appealing alternative for climatological hail studies. These instruments provide spatiotemporally uniform, three-dimensional, direct measurements of hydrometeors, making them ideal for climatological studies, assuming sufficient archive length and spatial coverage \citep{Saltikoff19}. When restricted to single-polarized radar data (as is the case in Australia), the Maximum Estimate Size of Hail \citep[MESH,][]{Witt98} parameter emerges as perhaps the most thoroughly validated hail diagnostic. Although MESH is not a reliable predictor of hail sizes \citep{Wilson09, Ortega18}, it has been shown to usefully discriminate hail occurrence within thunderstorms \citep{Cintineo12, Ortega18, Murillo19}. MESH has also been used to generate numerous regional hail climatologies in Europe \citep[\eg][]{Kunz15, Lukach17} and the United States \citep{Cintineo12, Wendt21}. The latter were created using the US NEXRAD radar archive, which represents the ``gold standard'' for radar networks from a climatological standpoint, consisting of over 160 identical radars operating over three decades. In contrast, the Australian radar network is highly diverse in terms of archive length, beamwidth, frequency, and receiver type \citep{Level1b}, and covers only $\sim$30\% of the country's total land surface. To date, these factors have limited radar-based hail studies to simple diurnal/seasonal analyses in 10 locations \citep{Dowdy20}, or more detailed regional climatologies for 18 years in Brisbane \citep{Soderholm17}  and eight years in Brisbane and Sydney \citep{Warren20}. 

In this study, our objective is to expand on previous work by creating a national hail climatology. To achieve this goal, we detail how we must address the unique challenges posed by the Australian radar archive, including normalizing differences in radar frequency and archive durations, and interpolating between large regions with no observations. These methodological elements are outlined in Sections \ref{s:methods} (methodology) and \ref{s:freq} (radar frequency correction). We begin our analysis at a continental scale in Section \ref{s:national}, by generating Australia's first national radar-based hail climatology. We then identify five regions of interest within the national climatology to investigate on a regional scale in Section \ref{s:regional}, providing a more comprehensive analysis for each. Finally, we summarize our findings and suggest directions for future work in Section \ref{s:conclusion}.


\section{Methodology}\label{s:methods}

\subsection{Data Processing}\label{proc}

\begin{figure*}[t]
\centering
\includegraphics[width=0.9\textwidth]{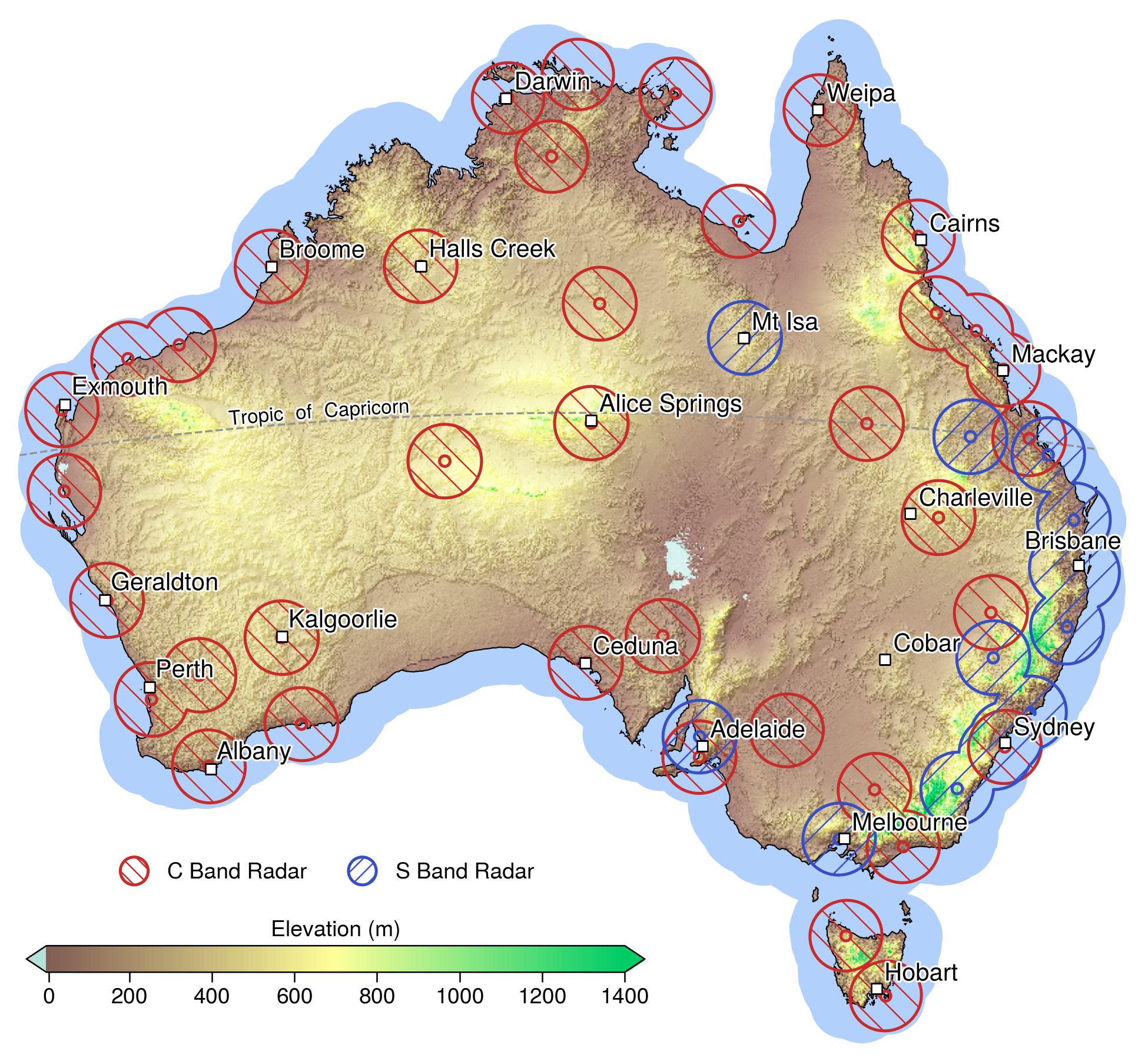}\\
 \caption{\label{archive} An illustration of the radar network used in this study. Contours demonstrate the spatial data coverage of the radar network (ranges 20--140 km), with C- and S-band radars in red and blue, respectively. The ocean within 140 km of the mainland, excluding islands other than Tasmania, is filled with blue, and background shading illustrates topography.}
\end{figure*}

The radar data used in this study are obtained from Level 1b of the Operational Radar Dataset in the Australian Unified Radar Archive \citep[AURA,][]{Level1b}. Our analysis encompasses all available data between 2000--2022, and mandates that each radar must contain over five years of operational observations in order to be included. Certain radars have also been excluded due to data quality concerns, such as geographically remote radars that did not historically transmit volumetric data, radars with persistent beam blockages, and various airport radars that are unsuitable for meteorological applications. These restrictions narrow our analysis to 52 radars, as illustrated in Fig.\ \ref{archive}. Among the 52 radars examined, only eight are polarimetric\footnote{The Bureau of Meteorology's Radar and Observation Network Uplift program dictates that all newly installed radars are polarimetric. At the time of writing, 16 of the 66 current operational radars are polarimetric, with this number expected to rise in years to come.} and fourteen are S-bands (wavelengths $\sim$10 cm); both of which have important implications for our study, and merit separate discussions in Sections \ref{s:methods}\ref{hail} and \ref{s:freq}, respectively. The Level 1b dataset contains quality controlled radar data in polar coordinates, and due to the single-polarization limitations for most of the archive, the horizontally polarized equivalent radar reflectivity factor ($Z$, hereafter reflectivity) is used exclusively for hail detection. A filtering procedure is applied to remove ground clutter contamination \citep{Gabella02, Heistermann13}, and radar reflectivity is calibrated using a ground clutter monitoring and spaceborne radar comparison technique \citep[refer to][for additional details]{Louf23}. The latter is particularly important for our purposes, as calibration errors are known to severely impact hail size estimates \citep[\eg][]{Warren20}. Recent three-way calibration checks for the Australian archive \citep[between ground-based, ship-based and spaceborne radars,][]{Protat22} have found that post-correction calibration errors fall within $\pm$0.5 dB, theoretically leading to acceptable MESH errors of less than 5\% \citep[][their Appendix A.]{Warren20}.  

Special consideration must be given for data quality issues when using radar data for climatological studies, particularly when the climatological variable in question may be strongly affected by weak nonmeteorological echoes, such as chaff, anomalous propagation and sidelobe contamination \citep[\eg][]{Fabry17}. These effects are unlikely to impact our results, given the relative echo strength and the elevated nature of the observations used to derive MESH (>3 km altitude). A possible exception may arise when spurious high-altitude, high-reflectivity echoes are detected due to ground clutter from elevated topography, especially during atmospheric conditions conducive to anomalous propagation. We contend that these cases occur infrequently in the Australian network due to the absence of prominent topographic features (highest peaks $\sim$1500 m, Fig.\ \ref{archive}), and direct the reader to the geometric justification given in Section 3.5 of \citet{Nisi18} for a highly mountainous region. As in previous studies \citep[\eg][]{Cintineo12}, a manual quality control procedure was applied to all identified hailstorms in the archive, which eliminated 1342 individual hail detections deemed spurious by a trained observer. We have determined that these evidently erroneous radar returns (such as complete volume saturation of >100 dBZ reflectivity) are likely attributable to hardware or signal processing failures and predominantly occur in the earlier portions of the archive for older radars. 

The final data processing consideration involves the amalgamation of overlapping reflectivity observations from individual radars. Previous studies have highlighted the importance of integrating information from multiple radars in order to fill gaps in radar scanning strategies and compensate for range dependencies \citep{Stumpf04, Ortega06}. However, recent research by \citet{Brook22} showed that gridding methods used to merge data into mosaics can degrade data quality, primarily through over-smoothing in the case of weighted average methods \citep[\eg][]{Langston07}. Nearest-neighbor and linear interpolation methods can also propagate observational noise and introduce first-order discontinuities into the analysis \citep[\eg][]{Lakshmanan06}. Single-polarized hail detection methods are highly dependent on the vertical distribution of reflectivity columns, the reconstruction of which is considerably degraded by both of the aforementioned gridding artifacts. Consequently, we have opted to regrid reflectivity observations using the variational interpolation method outlined by \citet{Brook22} for each radar separately in our study, with a grid spacing of 1 km horizontally and 500 m vertically. This approach allows for quality control of individual radars without removing or recalculating the resulting mosaic when erroneous data from one radar is removed and ensures that isolated radars with no overlapping coverage are processed in the same manner as those that overlap. However, it also requires an additional post-processing step to merge the resulting hail detections for overlapping radars, which is detailed in Section \ref{s:methods}\ref{hail}. We deemed the regridding process essential in order to capitalize on its superior characterization of deep convective plumes, despite the considerable computational resources required for applying the variational approach to all hail events in the archive. These methodological improvements ensure that the gridded reflectivities are closest to those at the native resolution of the radar, where the original MESH constants were defined \citep{Witt98}.

\subsection{Hail Identification}\label{hail}

\begin{table*}[t]
\begin{center}
\begin{tabular}{p{0.3\linewidth}p{0.3\linewidth}c}
\topline
Study & Hail Metric & Threshold (mm)\\
\midline
\citet{Wilson09} & Hail reports $\geq19$ mm & 28\\
\citet{Cintineo12} & Hail reports $\geq19$ mm & 29\\
\citet{Ortega18} & Hail reports $\geq25.4$ mm & 30\\
\citet{Murillo19} & Hail reports $\geq25.4$ mm & 29\\
\citet{Warren20} & Insured hail damage & 32\\

\botline
\end{tabular}
\end{center}
\caption{\label{t1}MESH thresholds used to identify severe/damaging hail in various studies. Refer to each study for further details on the MESH matching/thresholding procedure used therein.}
\end{table*}

The Maximum Expected Size of Hail parameter \citep[MESH,][]{Witt98} is used for hail detection in this study. Although originally developed as an object-based, single-valued parameter, we adopt the common approach of calculating MESH on regular grids \citep[\eg][]{Stumpf04}. This process begins with the calculation of the kinetic energy flux based on the semiempirical relationship first introduced by \citet{Waldvogel78I},

\begin{equation} \label{hke}
    \dot{E} = 5.0 \times 10^{-6}\cdot 10^{0.084Z} ,
\end{equation}

\noindent
where $Z$ is in logarithmic units (dBZ). The hail kinetic energy flux is subsequently integrated across height levels that are cold enough ($<0^\circ$C) and possess reflectivities large enough ($>40$ dBZ) to likely result from hail. Formally, these constraints are implemented through piecewise linear weighting functions for both conditions ($W(T)$ and $W(Z)$, respectively). The weighted, vertical integration of hail kinetic energy flux results in a parameter termed the severe hail index (SHI), which is formalized as follows, 

\begin{equation}
    \text{SHI} = \frac{1}{10}\int_{H_0}^{H_{\text{top}}}W(Z)\cdot W(T)\cdot \dot{E}\cdot dH,
\end{equation}

\noindent 
where $H_0$ is the height of the 0$^\circ$C isotherm and $H_{\text{top}}$ is the echo-top height. Finally, \citet{Witt98} established the MESH formulation by empirically fitting SHI values to the 75th percentile of 147 hail report sizes, yielding MESH $= 2.54(\text{SHI})^{0.5}$. \citet{Murillo19} recently refit the SHI--MESH relationship using more than 5000 hail reports; however, these updated constants are not used here due to the alternate radar processing methods used in that study\footnote{\citet{Murillo19} used a multi-radar reflectivity mosaic, smoothed using weighted average interpolation \citep[GridRad,][]{Homeyer17b}. MESH estimates are extremely sensitive to reflectivity biases \citep{Warren20}, meaning the SHI--MESH relationship derived for smoothed reflectivities is not suitable for the high-resolution grids used here \citep{Brook22}}. MESH has proven to be an effective hail discriminator; however, the exact MESH threshold used to detect hail varies slightly in the literature. Threshold values were defined to optimally identify severe/damaging hail, and several relevant studies and their corresponding MESH thresholds are given in Table \ref{t1}. Given the similarity among these five preceding studies and the aforementioned $\sim$5\% MESH errors associated with radar calibration, we deem it appropriate to take the average across them, which, when rounded to the nearest millimeter, results in a threshold of 30 mm. 

We address the discrete temporal sampling rate of weather radars by implementing an advection correction algorithm from the PySTEPS package \citep{Pulkkinen19}. This method interpolates MESH values along storm tracks by shifting values at time $t$ forward and values at time $t+1$ backward, before linearly weighting the two fields according to their temporal proximity \citep{Anagnostou99}. Advection velocities are calculated across the domain by applying the Lucas--Kanade optical flow method to the MESH fields. The procedure is applied over contiguous time periods with valid hail detections (volumes with $\geq30$ mm MESH), and temporal interpolation is performed at minute intervals. These time periods are calculated such that hail detections separated by less than or equal to 30 minutes are grouped together. This threshold was selected to accommodate earlier years of the archive that had 10 minute (rather than five minute) volume sampling times, ensuring hail streaks\footnote{We follow the terminology outlined in \citet{Changon70} by referring to a \textit{hail streak} as ``an area of continuous hail with temporal coherence, which is considered an entity of hail generated within a thunderstorm''. A combination of spatiotemporally proximal hail streaks may then be subjectively grouped to form a \textit{hail swath} for an event.} from a single storm are not artificially split in cases with missing volumes due to hardware or communication failures.

\begin{figure*}[t]
\centering
\includegraphics[width=1\textwidth]{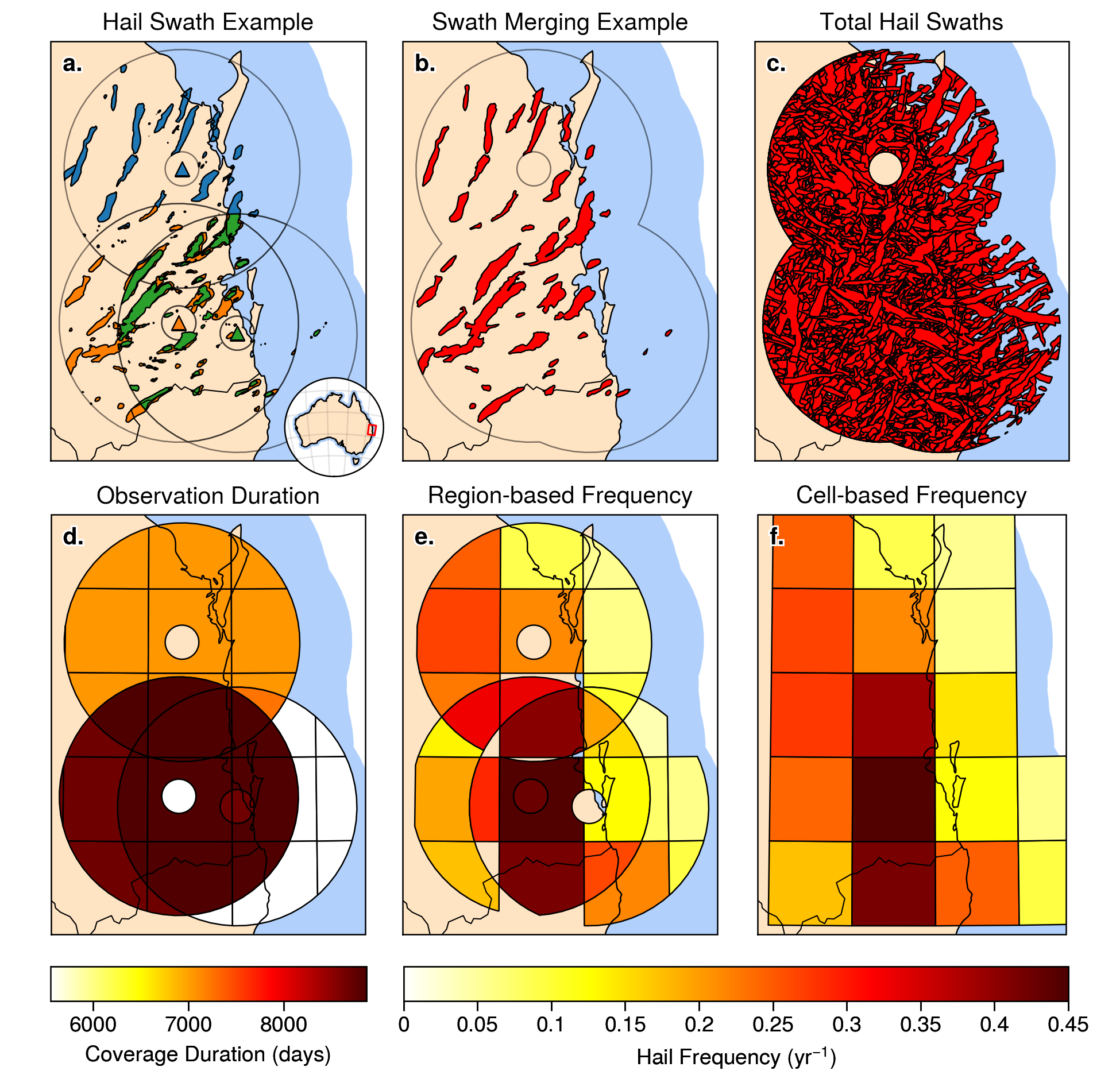}\\
 \caption{\label{explain} An illustration of the methods used to calculate the spatially invariant hail climatology. (a) Hail swaths outlined by the 30-mm MESH contour for the Mt. Stapylton (green), Marburg (orange), and Gympie (blue) radars on the 16th of December 2010, (b) merged hail swaths for the same case study (red), and (c) merged hail swaths for the entire archive. (d) The number of days observed for each coverage region at 100 km grid spacing, (e) annual hail frequencies for each coverage region, and (f) cell-based hail frequencies for a uniform grid.}
\end{figure*}

Our analysis then departs from conventional MESH studies by adopting an object-oriented approach, where individual hail streaks are created by contouring interpolated MESH grids at the 30-mm level. Each hail streak is defined by the 30-mm contour polygon, and an example hail swath composed of hail streaks from three overlapping radars is shown in Fig.\ \ref{explain}a. We conservatively \citep[relative to prior studies, \eg][]{Soderholm17, Warren20, Cintineo12} limit swaths to a radial range of 20--140 km in our analysis to mitigate an underestimation of the true hail frequency due the radar's scanning geometry (cone of silence and beam broadening). This conservatism is justified for the Australian radar archive, given the varying beam widths (between 1--2$^\circ$, refer to Table C1) and the dual-PRF artifacts present beyond 140 km. A preliminary analysis of the range dependency of hail frequencies (not shown here) revealed that hail frequencies were relatively stable between our chosen range thresholds and no obvious underestimation was observed for radars with larger beam widths. We direct the reader to \citet{Bunkers13} for a more comprehensive discussion of these effects. 

Hail streaks from individual radars are then merged by taking the geometric union of spatiotemporally overlapping polygons (refer to Figs.\ \ref{explain}a and \ref{explain}b), yielding spatially uniform, merged hail swaths. Merged hail streaks are assigned a range of statistics, including their area, start and end times, and optical flow velocities. Most of these statistics require an object-based approach, as they may not be easily calculated from an accumulated, gridded field \citep{Nisi18}. The resulting merged hail archive (\eg Fig.\ \ref{explain}c), along with the associated object-based statistics, form the basis of our national and regional hail climatologies. 

\subsection{Frequency Normalisation}\label{norm}

 The precise areal definition of hail streaks in our object-based approach permits a novel, resolution-invariant method for calculating yearly or monthly hail frequencies. In a traditional approach using temporally accumulated (\eg daily) MESH grids, hail frequencies can be evaluated at the original grid resolution by counting the number of times each pixel exceeds a mesh threshold \citep[\eg][]{Cintineo12}. However, by chance, some isolated grid points may not experience a single hail storm throughout the archive duration, leading to the spurious claim that the true climatological frequency at that point is zero. One may attempt to account for the finite archive duration by applying various ad hoc smoothing techniques \citep[\eg][]{Cintineo12, Murillo21}, or calculating hail frequencies on a coarser spatial grid. Hail frequencies calculated using the latter approach \citep[\eg on a 10 km grid in][]{Warren20} yield higher values, as the number of hail detections within a 100 km$^2$ area is predictably greater than the number at each individual grid point. This resolution bias in hail frequencies prohibits reliable comparisons between studies with different grid sizes [\eg \citet{Soderholm17} and \citet{Warren20}], and confuses the true meaning of hail frequency estimates. 
 
 In this study, we define hail frequency as the number of times hail is expected at any single point within a grid box, per normalization period (\eg yearly or monthly). We justify this choice as more intuitive to an observer, as it attempts to answer the common question: ``How many times per year are you likely to observe hail?'' This is distinct from: ``How many times per year will it hail within 10 km from here?'' This spatially invariant hail frequency is calculated by intersecting all hail streaks with a spatial region defined at any resolution. The cumulative area of all hail streaks within that region is then normalized by the area of the region ($A$) and its radar coverage duration ($N$, with quality controlled days removed). Formally, hail frequencies ($f_h$) are defined for each region as follows,

\begin{equation} \label{freq_norm}
    f_h =  \frac{365.25}{N} \times \frac{\sum_{i=0}^{n}A_i}{A}
\end{equation}

\noindent where $A_i$ is the area of the intersection between the i$^\text{th}$ of $n$ hail streaks and the analysis region. Note that for monthly frequency calculations, 365.25 is replaced by 365.25 / 12 = 30.4375, and $N$ and $A_i$ only count days and hail streaks that occur within that month\footnote{Also note that ``normalized months'' (\ie twelve months spaced evenly across the year) are used in favor of calendar months to account for the effects of varying month lengths.}. This approach can also be applied to gridded hail climatologies by counting the number of MESH grid values that surpass the threshold, before normalizing by the total number of grid points within the coarser climatology grid.

The irregular shape of overlapping radar coverage domains (see, for instance, Fig.\ \ref{explain}a for three radars) further complicates the specification of hail frequencies in a regular grid. In grid boxes with overlaps in coverage, it is difficult to prescribe an $N$ value, as different regions within the grid box have been observed for different periods of time. This is especially true for larger grid boxes, such as the 100$\times$100 km grids illustrated in Fig.\ \ref{explain}d. The solution is to intersect the radar coverage geometries with the underlying climatology grid, producing irregular ``coverage regions'', each with a unique coverage duration. Hail swaths are then intersected with the coverage regions (black lines in Figs.\ \ref{explain}d-f), and an individual hail frequency is calculated for each coverage region according to Equation \ref{freq_norm}. The result of these region-based frequency calculations is illustrated in Fig.\ \ref{explain}e. Finally, we aggregate all region-based frequencies within each grid cell by preferentially weighting those regions with larger areas, longer archive durations, and more overlapping radars ($R$), as these factors all reduce uncertainty in the underlying frequency estimates. Weighted averages are computed as follows,

\begin{equation}
    f_h = \frac{\sum_{j=0}^{m}\left(f_{h,j} \cdot A_j N_j R_j\right)}{\sum_{j=0}^{m} A_j N_j R_j}
\end{equation}

\noindent where subscripts denote values for the $j^\text{th}$ of $m$ coverage regions within each grid cell. In this manner, we are able to produce resolution-invariant gridded hail frequencies.

\subsection{Variational Interpolation}\label{interp_methods}

The Australian radar network covers roughly 30\% of the country's land surface, the distribution of which is illustrated in Fig.\ \ref{archive}. As a result, some form of interpolation is required to create a continuous hail climatology throughout the country. This interpolation problem is particularly challenging due to large data voids, local variability in hail frequencies, and data noise introduced by the finite sampling period of the archive. Variational inverse methods are well suited to such a problem, as certain ``prior knowledge'' may be formally introduced to regularize an initially ill-posed system \citep[\eg][]{Barth14}. Here, we seek to establish the following assumptions to help constrain the problem at hand:

\begin{enumerate}
    \item Radar-derived hail frequencies are a noisy but representative sample of the true climatological values.
    \item The true climatological hail frequency varies smoothly in space and time at our chosen resolution.
    \item No local extrema lie within data voids, and gaps are sufficiently constrained by the surrounding data.
    \item Hail frequencies reduce to zero at some distance from the coast.
\end{enumerate}

We will briefly discuss the reasoning behind each of these assumptions before outlining their mathematical implementation. First, we justify the representativeness of our frequency estimates in Assumption 1 by noting the relative length of our archive (average $\sim$17 years, Appendix B) compared with seminal radar-based climatologies [\eg four years in \citet{Cintineo12} and eight years in \citet{Warren20}]. We choose suitably sparse grid spacing to satisfy assumption 2 (100 km), which theoretically dampens high frequencies in the input data through spatial averaging and permits the use of smoothing constraints to ``spread'' valid information into data voids. Prior studies of the hail climatology in Australia \citep[eg][]{Allen16, Bedka18b, Prein18}, provide justification for assumptions 3 and 4, where hail hotspots are not expected outside the current radar coverage or significantly out to sea. The extent to which the available data constrains the analysis within data voids will be investigated more thoroughly below. 

We choose to perform the interpolation iteratively at two spatial scales: first at 200 km resolution to produce a provisional field, then at the final 100 km resolution using the former as a background field ($\varphi_b$). This type of nested analysis is common in variational problems \citep[\eg][]{Laroche94, Dahl19}, and is useful in our case as valid data spreads more readily at a coarser resolution, providing a useful background for the final analysis. The cost function used in both cases is as follows,

\begin{equation} \label{cost}
    \displaystyle{\min_{\varphi}} \left\{\left|\left|f_h - \mathcal{R} \varphi\right|\right|_2^2  + J_s(\varphi) + J_d(\varphi) + J_{b}(\varphi,\varphi_b)\right\},
\end{equation}

\noindent where $\varphi$ is the interpolated hail frequency field, $\mathcal{R}$ is a restriction operator that samples the analysis field at data locations and the standard $\ell_2$ norm notation is used throughout the text $||x||^2_2 = \sum_{i=1}^Nx_i^2$. The second term in the cost function is a second-order smoothing constraint, implemented as,

\begin{equation} \label{js}
    J_s(\varphi) =  \lambda_{h} \left(||\varphi_{yy}||_2^2 + ||\varphi_{xx}||_2^2\right),
\end{equation}

\noindent 
where $\lambda_h$ is a weighting constant and double subscripts denote second-order partial derivatives, \eg $\varphi_{xx} = \pdv[2]{\varphi}{x}$. Second-order smoothing constraints lead to a visually pleasing minimum curvature solution, and the 3-point, centered, finite difference stencil helps spread valid data into data voids \citep[][]{Briggs74, Shapiro09}. 

Although $J_s$ may partially remove speckle noise from the frequency data, \citet{Brook22} found that this constraint alone is unsuitable for denoising due to unwanted by-products produced when heavily weighting this constraint \citep[oversmoothing and extraneous inflections in data voids,][]{Shapiro10b, Brook22}. Complementary inclusion of an additional anisotropic total variation denoising constraint \citep{Rudin92} was found to adequately denoise the data, while damping extraneous inflections within the data voids and retaining high amplitude peaks within the analysis fields \citep[refer to Appendix B in][]{Brook22}. The denoising constraint is as follows, 

\begin{equation} \label{jd}
    J_d(\varphi) = \lambda_d \left(||\varphi_y||_1 + ||\varphi_x||_1 \right)
\end{equation}

\noindent where $\lambda_d$ is a weighting constant, subscripts indicate first-order partial derivatives (\eg $\varphi_{x} = \pdv{\varphi}{x}$), and the $\ell_1$ norm notation denotes $||x||_1 = \sum_{i=1}^N|x_i|$. Forward difference derivatives are used by convention \citep{Rudin92}.

Finally, the background penalty helps constrain regions outside the data coverage by penalizing deviations from an imposed background field ($\varphi_b$). At the provisional resolution (200 km), the background field is uniformly zero and only applies to grid points over the ocean through the application of a weighting matrix that increases exponentially from 0 over land to 1 out to sea ($W_b = \exp{-r^2/d^2}$) where $d$ is the distance from the coast and $r$ = 400 km controls the rate of weight increase). At the final resolution, the provisional retrieval is used as the background field, and $W_b$ is then set to 1 for grid points over land with no valid data. Formally, the background constraint is implemented as,

\begin{equation} \label{jb}
    J_b(\varphi, \varphi_b) = \lambda_b\left|\left|W_b\left(\varphi-\varphi_b\right)\right|\right|_2^2 
\end{equation}

\noindent
where $\lambda_b$ is once again a weighting constant. This constraint serves a dual purpose: it constraints the analysis to zero around the borders of the domain to prevent any extraneous inflections or boundary effects, and assists in filling large data voids with values from the provisional analysis, permitting the use of less aggressive smoothing constraints to prevent oversmoothing.

The aforementioned data voids may lead some readers to question the extent to which the surrounding data actually constrains the interpolated values within. In an attempt to satisfy these concerns and justify assumption 3,  we apply the “clever poor man’s error” technique, which is an efficient method for assessing data coverage in variational interpolation problems \citep{Troupin12, Beckers14}. The procedure takes place as follows: set all available data to unit values, reduce the weighting constants by a context-specific factor, and perform the analysis as normal. Due to the difficulties involved in prescribing an exact covariance function for our interpolation procedure, we assume Gaussian covariance and consequently reduce the weights in the second step by a factor of $\sqrt{2}$ \citep{Beckers14}. The resulting field approximates the ``error reduction factor'' (\ie the extent to which the available data reduces analysis errors below background error state), where values close to one indicate the problem is well constrained by the data. 

We treat the provision of weighting constants in Equations \ref{js}--\ref{jb} as a parameter tuning problem \citep[\eg][]{Brook22, Brook23}. Suitable weights are derived experimentally and ultimately determine the relative strength of each constraint. The chosen values for our interpolation problem are as follows: $\lambda_h = 0.5$, $\lambda_d = 0.01$, and $\lambda_b = 2$. The split Bregman optimization algorithm is used to minimize Equation \ref{cost} due to its applicability for mixed $\ell_1$--$\ell_2$ norms \citep[limited to 5 inner, and 10 outer iterations;][]{Goldstein09, Ravasi20}.

\section{Radar Frequency Correction} \label{s:freq}

\begin{figure*}[t]
\centering
\includegraphics[width=.99\textwidth]{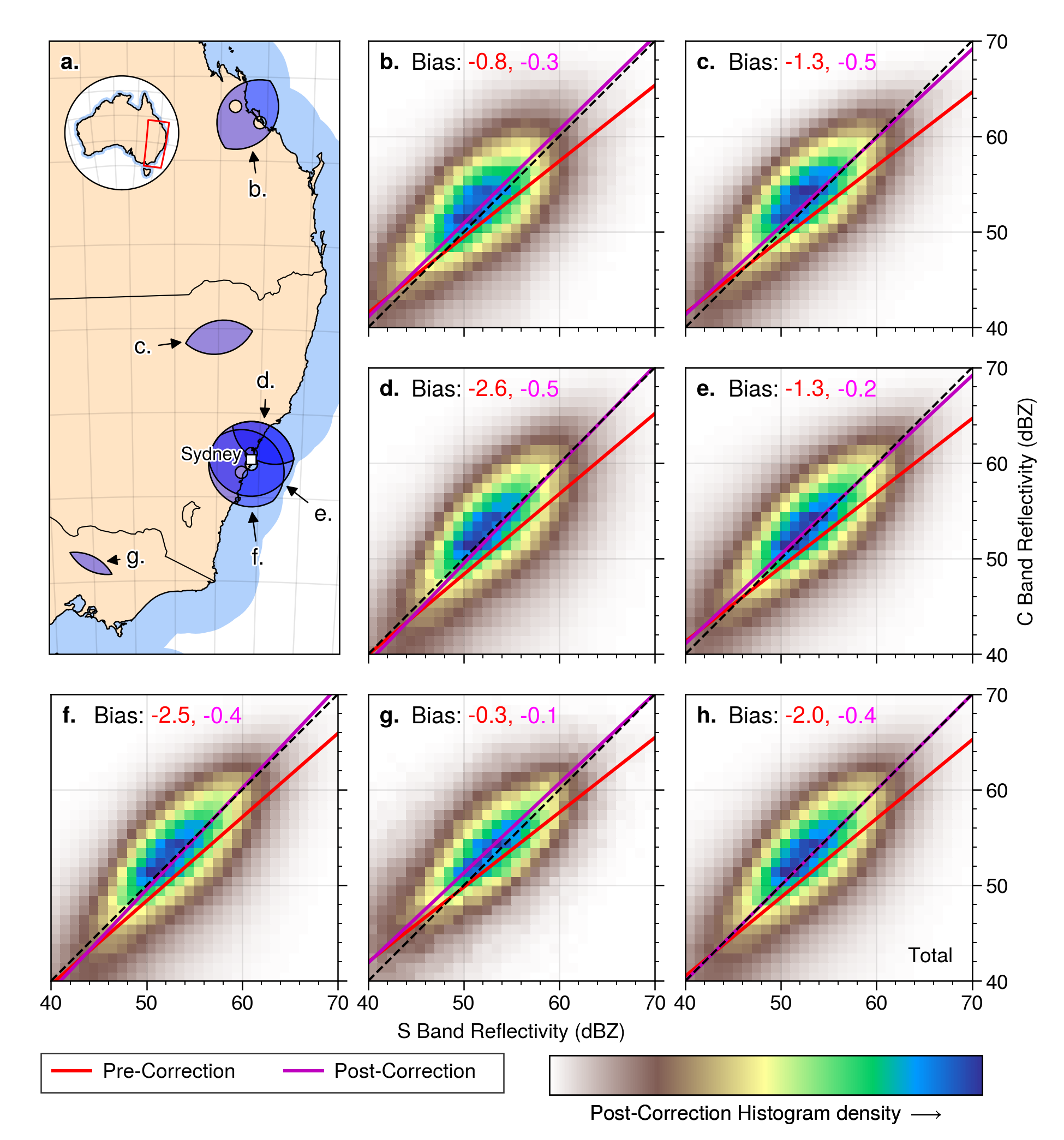}\\
 \caption{\label{freq_pic} An illustration of the hail reflectivity fitting process employed in this study. (a) Selected C- and S-band radar overlaps on the East coast of Australia, with labels indicating the panel in which the resultant data are shown. (b)-(g) Data fits for each region, including the precorrection fit between C- and S-band hail reflectivity (red), the corresponding postcorrection fit (magenta), and annotations showing the respective fit biases. A 2D histogram for the post-fit reflectivity data is included in each panel, shaded by qualitatively higher data densities. (h) The same as for regional overlap panels, except for all data combined.}
\end{figure*}

A major limitation of the approach outlined in Section \ref{s:methods} is the utilization of C- and S-band radars in the Australian archive (wavelengths $\sim$5 and $\sim$10 cm, respectively). MESH was developed using only S-band radars from the US WSR-88D radar network, and several theoretical challenges arise when attempting to apply it to C-band radars. One primary concern is the increased signal attenuation due to hail at C-band \citep{Battan71}, causing the observed reflectivity (and subsequent MESH values) to diminish relative to S-band. Various correction techniques have been proposed for attenuation at C-band \citep[\eg][]{Gu11}, however they require polarimetric data and undercorrect hail-induced attenuation \citep{Borowska11}. To the authors' knowledge, no single-polarized hail attenuation correction exists and thus none is implemented in this study. Instead, we aim to indirectly account for attenuation in a climatological sense using the corrections that follow\footnote{\label{note2}We differentiate between a correction made on a climatological basis (which aims to offset the net climatological difference between C- and S-band radars) and one that aims to correct attenuation effects for individual cases.}, while acknowledging the inevitable loss of information at C-band. 

The second major methodological challenge in implementing MESH at C-band arises due to non-Rayleigh scattering. Both theoretical scattering calculations \citep[\eg][]{Ryzhkov13a} and observational studies \citep{Kaltenboeck13} indicate that complex scattering effects from hail are more prominent at C-band, reducing the reflectivity response relative to S-band. Recently, \citet{Cecchini22} refit the original $Z$--$\dot{E}$ relationship \citep[Equation \ref{hke},][hereafter W78]{Waldvogel78I} at both C- and S-bands using updated mass/terminal velocity relationships, while also considering the nonsphericity, varying density and complex scattering effects of natural hailstones. Their analysis confirmed that $\dot{E}$ is strongly modulated by the frequency of the illuminating radar, as C-band radars are less sensitive to large hail. However, since the MESH formulation and thresholds used for hail detection (see Table \ref{t1}) have been developed using the W78 relationship, these updated and likely more accurate relationships may not be directly applied in our analysis. This would require refitting the empirical SHI--MESH function and hail detection thresholds using hail reports \citep[\eg][]{Murillo19}, which is outside the scope of our study. Instead, we attempt to generate a C-band analogue for the W78 relationship, which aims to account for the aforementioned frequency biases at C-band while remaining compatible with the existing MESH formulation.\footnote{The empirical C-band analogue of the W78 relationship derived here should not be compared directly to the C-band $Z$--$\dot{E}$ relationship derived in \citet{Cecchini22} due to the aforementioned methodological additions in their study. Future work should aim to assess whether their theoretical improvements result in more accurate hail size estimates at both C- and S-band when compared to hail reports.} C-band MESH may be of lower quality due to the reduced information content; however, previous studies have shown that it remains a skillful hail discriminator \citep{Kunz15, Skripnikov14, Strzinar18}. The empirical correction procedure we apply in our study ensures that frequency-related biases are optimally mitigated between C- and S-band radars. 

To establish our empirical correction, we examine six regions in the Australian archive with overlapping C--S-band coverage (illustrated in Fig.\ \ref{freq_pic}a). These regions were selected from the set of all C--S-band overlaps (refer to Fig.\ \ref{archive}) due to their sufficient overlapping coverage area ($>$4000 km$^2$) and number of co-sampled hail events ($>$100). The fitting procedure involved extracting time pairs where the C- and S-band radars both observed MESH $>$10 mm within the overlapping observation region, separated by a delay of less than two minutes to ensure temporal proximity. The 10-mm MESH threshold was selected to include events that may be initially underestimated by the C-band radars but could be incorporated into the 30-mm hail archive after correction. Once cosampling times were identified, all reflectivity data above the 0$^{\circ}$C isotherm (\ie only reflectivities that contribute to MESH) were extracted for columns exceeding the 10-mm MESH threshold, and the nearest pixels were matched between C--S bands for analysis. The pre-correction reflectivity fits for each region are shown in red in Fig.\ \ref{freq_pic}b--g, confirming the theoretical considerations discussed earlier: C-bands underestimate hail reflectivities compared to S-bands. Encouraged by the aforementioned frequency-dependent fits developed by \citet{Cecchini22}, we attempt to derive a similar correction empirically by linearly fitting observed reflectivities from the two frequency bands in an attempt to normalize the hail size estimates between them. As we aim to apply a single correction for all C-band radars in the archive (Fig.\ \ref{freq_pic}h), we combine the C- and S-band reflectivities for all six analysis regions and apply an iterative orthogonal linear regression technique to arrive at the following corrections:

\begin{align}
    Z_\text{S} &= 1.113 \cdot Z_\text{C} - 3.929 \label{z-fit}\\
    \dot{E} &= 2.34 \times 10^{-6} \cdot 10 ^ {0.093Z_\text{C}},\label{e-fit}
\end{align}

\noindent where Equation \ref{z-fit} was substituted into into the original W78 hail kinetic energy relationship (Equation \ref{hke}) to derive its empirical C-band analogue in Equation \ref{e-fit}. The chosen fitting technique and additional methodological considerations are outlined in Appendix A. 

To illustrate the effectiveness of the reflectivity correction in Equation \ref{z-fit}, it is applied to C-band data in each of the overlapping C- and S-band coverage regions, resulting in the 2D histograms shaded in Figs.\ \ref{freq_pic}b--h. Visual inspection of these histograms, along with the post-correction fits (magenta lines), reveal that the correction improves the negative bias for C-band reflectivity for all regions. Quantitatively, the mean bias improved from -2.0 to -0.4 dB for the total data, with the corresponding statistics for each region annotated on each panel. For reference, the pre-correction bias of -2 dB corresponds to a $\sim$20\% underestimation of MESH values at C-band \citep{Warren20}. Substantial spatiotemporal overlaps in Sydney (Australia's largest city) mean that the Kurnell/Woollongong (Fig.\ \ref{freq_pic}e) and Kurnell/Terrey Hills (Fig.\ \ref{freq_pic}f) overlaps account for $\sim$41\% and $\sim$29\% of the 18 million individual reflectivity comparisons in the total fit (Fig.\ \ref{freq_pic}h). However, it is encouraging to note that the reflectivity correction works equally well for regions outside of Sydney that contributed fewer data points. Next, we apply this correction to all C-band radars in the archive, aiming to derive a frequency-agnostic hail climatology. 

\section{National Climatology}\label{s:national}

\subsection{Data Coverage}\label{coverage}

\begin{figure*}[t]
\centering
\includegraphics[width=1\textwidth]{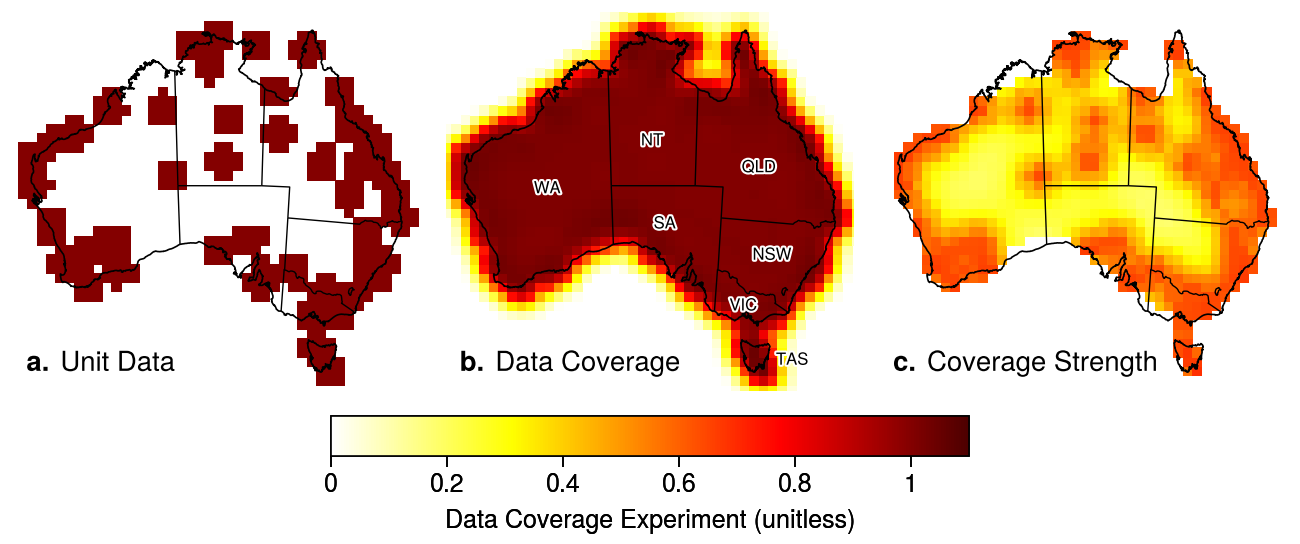}\\
 \caption{\label{interp_error} An illustration of important quantities for interpreting the variational interpolation method used in this study. (a) Data extent of the radar network (filled with unit data according to the poor man's error method). (b) The corresponding poor main's error estimate, a short-hand measure of data coverage. (c) The same but for ``strength'' of the available radar coverage, refer to text for more details.}
\end{figure*}

We begin our analysis by evaluating how well the available radar data constrains hail frequency estimates in Australia. Figure Fig.\ \ref{interp_error}a illustrates the radar network filled with unit data following the “poor man’s error” formulation (refer to Section \ref{s:methods}\ref{interp_methods}). Significant data voids are apparent throughout the country, particularly in central WA, northern SA and western NSW. These areas are devoid of radars due to their sparse population and the fact they largely encompass Australia’s deserts. To test whether our interpolation technique is capable of propagating valid information into these voids, the error reduction factor was calculated as outlined in Section \ref{s:methods}\ref{interp_methods} and shown in Fig.\ \ref{interp_error}b. In general, this quantity is equal to one over land and gradually decreases to zero away from the coastline, where the analysis is dominated by the background constraint and the radar data have minimal impact. The influence of the background constraint also extends slightly inland in regions with no radar coverage along the coastline (\eg northeast and southeast WA). Such regions are less constrained by the radar data and should be interpreted with this in mind. It is possible to adjust the variational tuning parameters to ensure error reduction factors are close to one everywhere; however, we found that such combinations are prone to oversmoothing. We justify our choice of parameters by noting that the poorly constrained regions are small and that the resulting interpolated fields are a faithful representation of the underlying radar data (\eg Fig.\ \ref{interp_freq}). Error reduction factors being close to unity almost everywhere over land validates our assumption that data voids are adequately constrained at our chosen analysis resolution. 

Although Fig.\ \ref{interp_error}b demonstrates that there is sufficient data coverage across the continent, the strength of that data coverage is highly heterogeneous. To quantify the strength of the data in regions with adequate coverage (error reduction factor > 0.9), we recalculate the error reduction factor for each grid point, this time setting the data at that grid point equal to zero. If the artificial data is placed in a data void, the analysis will be strongly influenced by it and the result will be close to zero. Conversely, if the point is inserted into a data-rich region, the addition of extra data will have less influence, and the result will be closer to one. The resulting field is therefore a quantitative assessment of the strength of radar coverage, with higher values indicating stronger coverage. 

Fig.\ \ref{interp_error}c reveals that the radar network strongly constrains the analysis around the coast, especially on the east coast. As noted earlier, the interior of the country contains large data voids, and the radar coverage strength is much lower there as a result. Interestingly, the coverage strength is not directly proportional to the distance to the nearest measurement. For example, small radar data gaps in southeast QLD and northeast NSW (Fig.\ \ref{interp_error}a) do not cause a significant drop in this quantity, as they are surrounded by valid data on all sides. In contrast, the coverage strength in western NSW drops immediately when crossing into that data void, as it is unbounded on its northern and western edges. Regions with weaker coverage strength should be assigned lower confidence, as the data is sourced from more distant measurements, and the analysis is more sensitive to measurement errors. The 0.25 coverage strength contour from Fig.\ \ref{interp_error}c is hatched in Figs. \ref{interp_freq}b and \ref{monthly} to illustrate the fact that both data coverage and coverage strength should be taken into account when interpreting the subsequent national climatology.

\subsection{Annual Hail Frequency} \label{annual}

\begin{figure*}[t]
\centering
\includegraphics[width=1\textwidth]{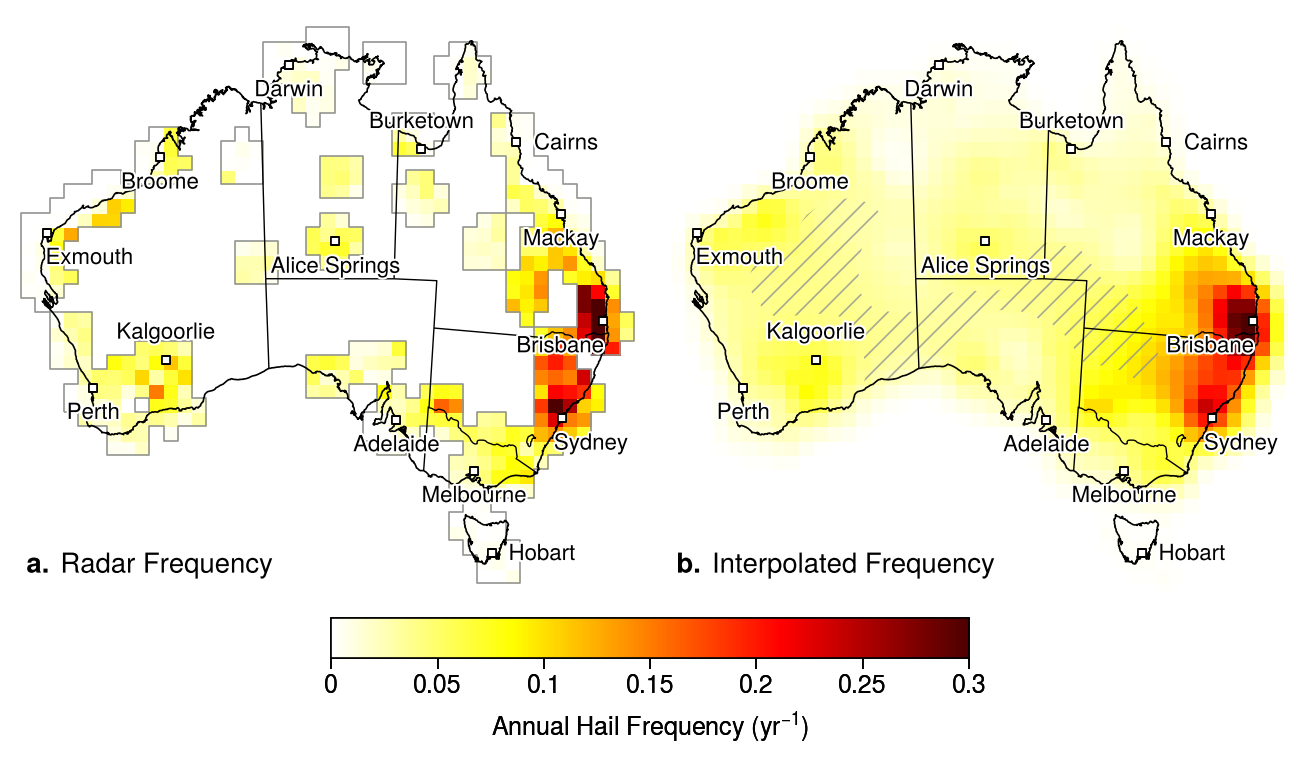}\\
 \caption{\label{interp_freq} A national radar-based hail climatology of Australia on a 100 km grid. (a) Climatological hail frequencies calculated using the Australian radar network, with gray outlines illustrating the data extent. (b) Nationally uniform hail frequencies calculated by interpolating the available radar data, with a coverage strength = 0.25 contour hatched in gray.}
\end{figure*}

Having discussed the intricacies of the chosen interpolation procedure, we present the resulting interpolated hail climatology in Fig.\ \ref{interp_freq}. The most notable feature is the broad hail frequency maximum located on the east coast of Australia, roughly situated south of Mackay in southeast Queensland and north of Sydney in northeast New South Wales. The highest hail frequencies are observed close to the coastline south of Brisbane, proximal to the Great Dividing Range (topography illustrated in Fig.\ \ref{archive}). These findings are consistent with previous studies in the Australian region \citep[\eg][]{Allen16, Bedka18b}. The position and size of this hail-prone region is similar to the hail environment-filtered overshooting top climatology in \citet{Bedka18b}, but extends slightly further inland than their study suggests. This inland propagation is partially influenced by a high number of observed events in southwestern NSW (near the SA/NSW/VIC border, Fig.\ \ref{interp_freq}a). Manual quality control affirms the legitimacy of these events, but the frequency disparity compared to the surrounding region suggests that this feature may be a climatological anomaly. Although the interpolation procedure smooths such features towards the ``climatological mean'', the anomalous information still propagates into the data void in western NSW. The limited data strength in this region (Fig.\ \ref{interp_error}c), limits our confidence in how far west this hail maximum extends. The recent installation of two radars (Hillston and Brewarrina) largely fills the western NSW radar void and will help alleviate these uncertainties as more data become available in years to come.

Radar-based hail frequencies drop off appreciably south of Sydney, with annual frequencies on the order of 0.05 year$^{-1}$ across Victoria and southern South Australia. Hailstorms in this region are known to be associated with the passage of frontal systems, which can initiate thunderstorms in otherwise marginal environments. \citet{Zhou21} demonstrated that these low instability and moderate to high shear hail environments are dominant in the high latitudes, where cooler temperature profiles also increase the likelihood of hailstones reaching the surface. Previous remote sensing studies have suggested that these relatively shallow hailstorms may result in an underestimation of hail frequencies in southern regions such as Melbourne and Canberra \citep{Bedka18b, Punge18}. While there are numerous hail reports in these areas \citep[\eg][]{Bedka18b}, the reported sizes are smaller on average, reflecting less hail growth due to reduced precipitable water and weaker instability \citep{Allen16}. As a result, a greater proportion of hailstorms and hail reports will be correctly overlooked by our 30-mm MESH threshold. It is also possible that MESH is poorly calibrated for detecting large hail due to these structural differences and underestimates the true frequency of hail in southern Australia \citep[having been empirically fit to hail size reports in Oklahoma and Florida,][]{Witt98}. Future work should investigate the relative accuracy of MESH estimates against hail reports for different high-population centers around the country. 

On the west coast, radar coverage is limited to the coastline in the north but extends further inland in the Wheatbelt region between Perth and Kalgoorlie. Both regions exhibit elevated hail frequencies ($>$0.1 yr$^{-1}$), the first along the coastline between Exmouth and Broome and the second in the Goldfields region around Kalgoorlie. Both signatures appear fragmented in the observed radar frequencies (Fig.\ \ref{interp_freq}a), with disparate frequencies detected at adjacent grid points. Once again, the interpolation procedure tempers isolated high-frequency points to estimate the climatological mean. The result is a broad region of elevated hail frequency from Broome to Exmouth and around Kalgoorlie. Notably, the latter does not extend to the coastline in the south near Perth, Australia's fourth most populous city. The land--sea contrast is especially apparent on the north coast of WA, where hail frequencies over the ocean are nearly zero, consistent with climatologically high values of convective inhibition (CIN) observed offshore in the region \citep{Riemann09}. Other minor hail frequency maxima are observed by radars in central Australia near Alice Springs, and in northwest Queensland. The former region coincides with a collection of hail reports in the Australian Severe Storms Archive \citep{Allen16}, although it is not identified in previous studies involving other remote sensing instruments \citep[\eg][]{Dowdy14, Bang19}. \citet{Bedka18b} did identify a local hail maximum in the goldfields region near the WA/NT/SA border, however, this is not mirrored in the radar archive, with the Giles radar west of Alice Springs showing little hail activity. The other local maximum near Burketown is also observed in the lightning climatology record of \citet{Dowdy14}, and is probably associated with low-level convergence in the Queensland heat low \citep{Sturman96} and the onshore passage of cloud lines from the Gulf of Carpentaria \citep{Drosdowsky87}. 

Tropical radars north of Broome and Cairns (latitudes $\sim$18 and 17${^\circ}$S, respectively) exhibit minimal hail activity in our analysis, which is noteworthy for two reasons: 1) these findings may represent the first climatological application of MESH at such latitudes, and 2) previous hail climatology methodologies tend to overestimate hail occurrence in the tropics. These methods include satellite brightness temperatures \citep[\eg][]{Bang19}, overshooting tops \citep[\eg][]{Bedka18b} and hail/severe storm proxies \citep[\eg][]{Prein18}. All such methods are sensitive to tall, overshooting clouds in high-instability environments, which occur commonly in the absence of severe surface hail in the tropics. Several complementary factors are likely responsible for the absence of hail in the region. Firstly, vigorous warm-rain processes below the freezing level are thought to simultaneously limit the amount of supercooled liquid water and increase the number of competing hail embryos in the hail growth zone \citep{Cotton10}. Second, the lack of wind shear in the tropics \citep[\eg][]{Allen16, Zhou21} means that hail-producing storms are comparatively shorter-lived, as outflows from mature cells disrupt their own inflows \citep{May02}. These structural differences likely reduce the potential updraft residence time, which is critical for hail growth \citep[\eg][]{Kumjian21}. 

Both structural and microphysical limitations discussed here dictate that hail observed aloft is mostly small \cite[1--2 cm,][]{May02}, and likely melts before reaching the surface due to typical temperature profiles in the tropics \citep{Foote84}. To the contrary, \citet{Bang21} demonstrated that spaceborne Ku-band radar reflectivity profiles are mostly consistent across tropical and subtropical hailstorms identified using different passive microwave detection methods. We propose that this discrepancy arises due to complex scattering and attenuation at Ku band radar frequencies, which reduces the reflectivity response for larger hail sizes relative to S-band and limits the information content available for hail size discrimination \citep{Ryzhkov13a, Cao13}. A notable finding in Section \ref{s:freq} is the substantial impact of these effects at C-band frequencies, implying even more pronounced consequences at Ku-band. Additionally, the results presented here indicate that unlike Ku-band radars, ground-based S-band and corrected C-band radars effectively distinguish larger hydrometeors likely to reach the ground without melting, thereby minimizing the overestimation of hail frequencies in tropical regions. Admittedly, this analysis is limited to the available radar coverage, which excludes most of the Kimberley coast region in northwest Australia (see Fig.\ \ref{archive})\footnote{An important exception is the Broome radar that covers the Dampier Peninsula on the western fringe of the Kimberleys.}, where other remotely sensed hail \citep[\eg][]{Bang19} and severe storm climatologies \citep[\eg][]{Dowdy14} notably highlight as a hotspot. However, based on the findings in Section \ref{s:national}\ref{coverage}, we expect that the available radar data adequately constrains hail frequency estimates across tropical Australia. The ground-radar-based assessment of tropical hail frequency shown here appears realistic when compared to Severe Storm Archive hail reports \citep{Allen16} and aligns quantitatively with a 20-year climatology compiled from press reports by the Queensland Regional Meteorological Office \citep{Frisby67}, with the 1-in-10 year (0.1 yr$^{-1}$) and 1-in-20 year (0.05 yr$^{-1}$) isolines terminating around Mackay and Cairns, respectively. 

\subsection{Hail Seasonality}

\begin{figure*}[t]
\centering
\includegraphics[width=1\textwidth]{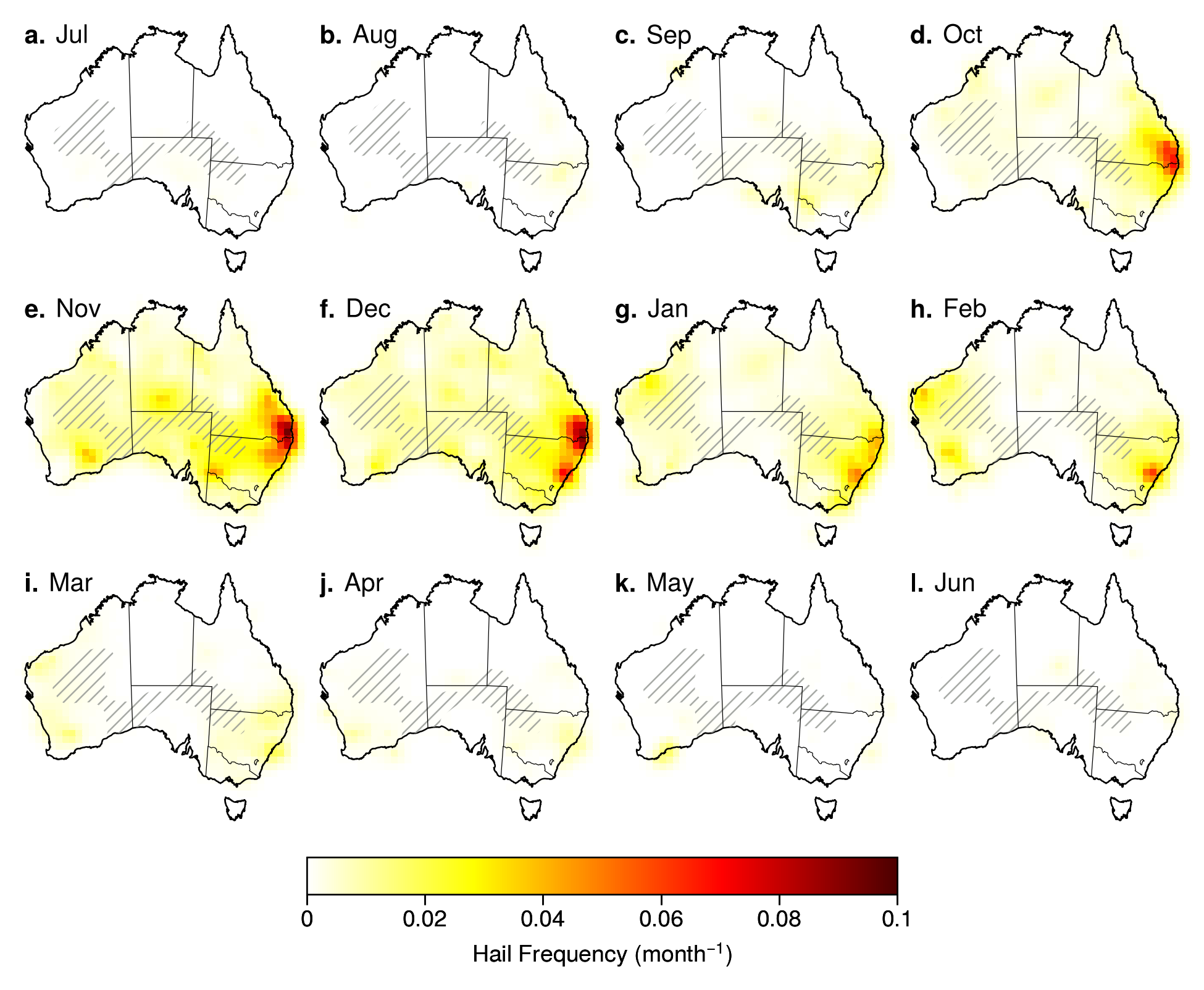}\\
 \caption{\label{monthly} Interpolated monthly hail frequencies derived from the Australian radar archive on a 100 km grid with a coverage strength = 0.25 contour hatched in gray. Months are normalized into twelve equally spaced periods and are displayed centered on the Austral summer.}
\end{figure*}

The final result presented in this section concerns the seasonality of the Australian hail climatology. Fig.\ \ref{monthly} presents interpolated monthly hail frequencies, calculated using normalized months to ensure an even sample size for each month. Segmenting the historical archive in this manner naturally limits the data available for each interpolation, potentially challenging our assumption that the observed data are a sufficiently representative sample of the actual climatological hail frequency. These limitations can lead to ``patchy'' hail frequencies for months such as November and February (Figs.\ \ref{monthly}e and \ref{monthly}h, respectively), which appear to be affected by climatological anomalies that arise due to the limited duration of the archive. We proceed in discussing the seasonal climatology with these limitations in mind. 

While some hail activity is observed on the east coast as early as September, the hail season begins in earnest in October. Early season events are mostly concentrated in southern Queensland and progressively move south as the season advances. Note that the gap observed between Brisbane and Sydney in December and January likely reflects the lack of radar observations in the region (Fig.\ \ref{interp_freq}a). By February, the East coast hail frequency maximum has shifted south to around Sydney, showing a steady poleward trend in the intervening months. This aligns with global satellite observations made by \citet{Zhou21}, who suggested that the poleward shift coincides with increases in temperature and moisture at higher latitudes throughout the summer. Locally, \citet{Allen14} and \citet{Warren20} showed how these effects lead to a poleward shift in instability as the Austral summer progresses, particularly along the east coast. Conversely, the weakening of lapse rates in the absence of favorable upper-level support results in an overall reduction in hail-producing instability/shear environments in Queensland as the summer progresses \citep{Allen14, Warren20}. As a result, the seasonal cycle is comparatively broader and later in NSW (November--March), a finding that is also reflected in hail reports from greater Sydney \citep[][]{Schuster05}.

Outside the east coast maximum, hail frequencies increase markedly throughout the country in November. This timing coincides with the arrival of continental heat troughs and their associated moisture transport to much of inland Australia \citep{Sturman96, Hung04}. This moisture, along with suitable upper-level support during late spring or early summer, provides favorable conditions for hail activity in those months, reminiscent of the seasonal climatology of convection in the Great Plains of the United States \citep[][]{Fabry17}. Under this hypothesis, the collective reduction in hail activity in January--February is a result of a warmer upper atmosphere as the summer progresses. \citet{Allen14} also cited the importance of inland troughs in mid to late summer for storm development across inland Western Australia (WA). Hailstorms on the west coast of the country remain subdued until later in the season around February, consistent with the overshooting top-based climatology of \citet{Bedka18b}. It is also worth noting that the only major hailstorm to impact the southwest capital city of Perth occurred similarly late in the season \citep[March 2010,][]{Buckley10}. Hailstorms persist in southern WA later than most of the country, likely due to their association with strong frontal systems that are capable of triggering hailstorms well into Autumn in the absence of strong instability \citep{Allen14}. 

\section{Regional Analysis}\label{s:regional}

\begin{figure*}[t]
\centering
\includegraphics[width=1\textwidth]{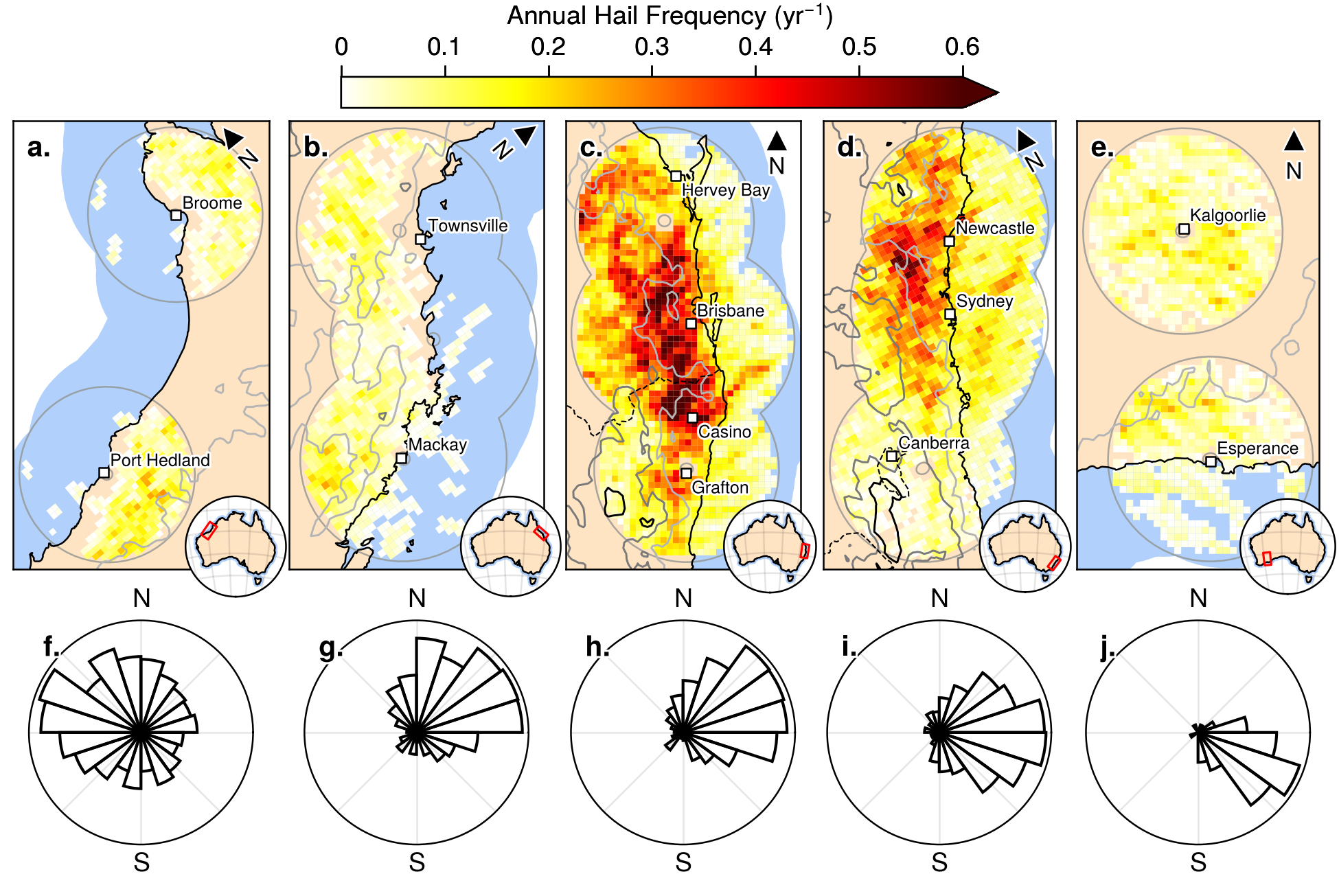}\\
 \caption{\label{maps} Regional hail climatologies from five regions of interest. (a)--(e) Hail frequencies direct from the radar data on a 10 km grid (no interpolation), for the northwest Australia (NWA), North Queensland (NQ), Brisbane, Sydney and southwest Australia (SWA) regions, respectively. Progressively darker gray contours show the 250, 750 and 1250 m topography levels, dotted lines represent state/territory borders and circular gray lines illustrate radar coverage extents. Note that panels (a), (b) and (d) have been rotated for ease of presentation, refer to north arrows and locations in inset panels for orientation. (f)--(j) Rose plots for storm directions in each region (pointing in the direction of storm motion).}
\end{figure*}

\begin{figure*}[t]
\centering
\includegraphics[width=0.9\textwidth]{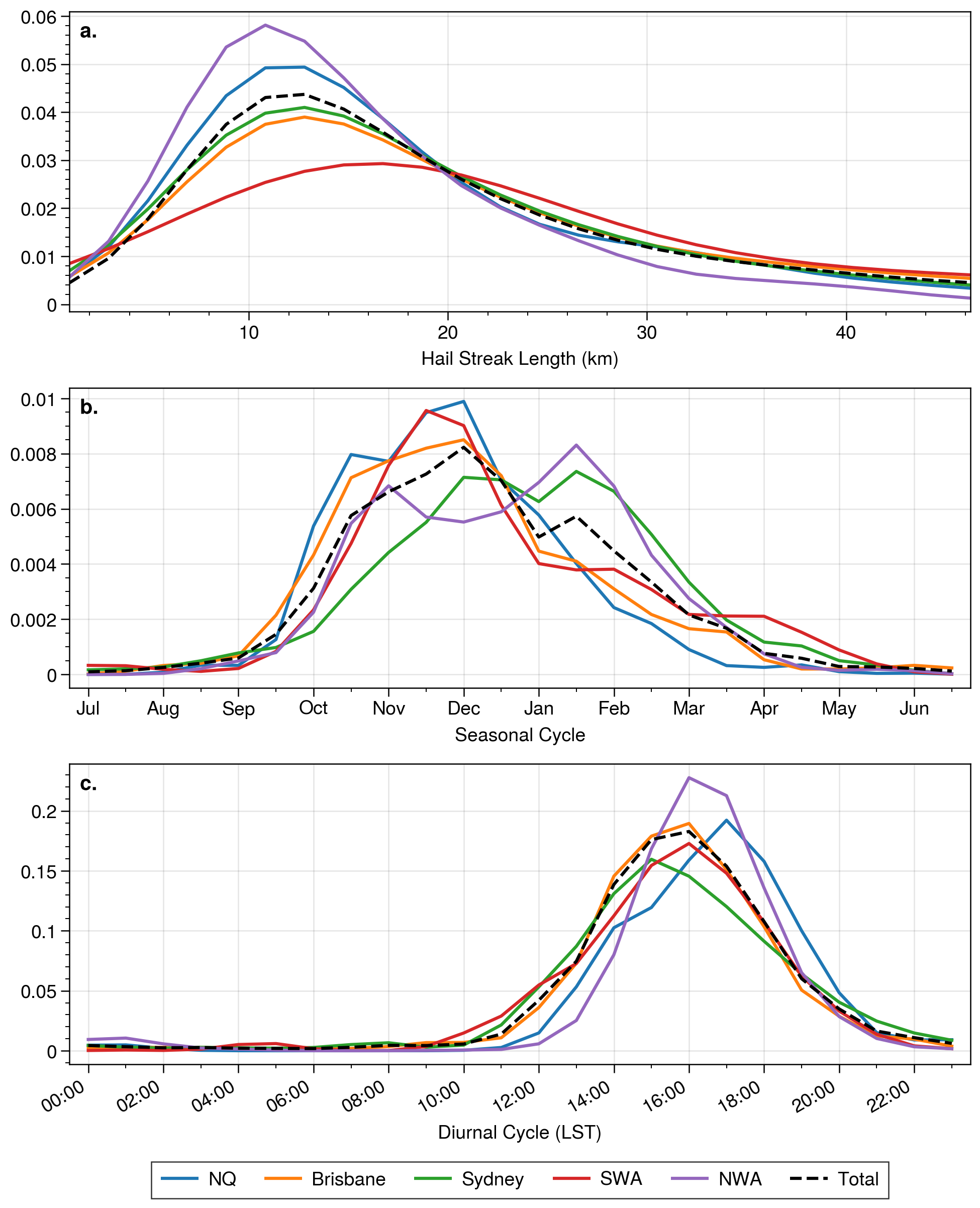}\\
 \caption{\label{stats} Normalized probability density functions of hailstorm statistics for the five regions of interest identified in Section \ref{s:regional} and the entire archive (total). (a) Length of the hail streak major axis, (b) seasonal cycle measured by hailstorm counts over 24 equally spaced time periods ($\sim$bi-weekly), (c) diurnal cycle measured by hourly counts of active hailstorms measured in local solar time (LST).}
\end{figure*}

Having presented the national climatology in Section \ref{s:national}, five regions of interest were identified for further investigation in this section. These regions were selected based on the discussion in Section \ref{s:national} (\eg northwest Australia and the Goldfields regions) and their significance as population centers (\eg Brisbane and Sydney). We have increased the spatial resolution of the regional climatologies to a 10 km grid; however, this is limited to areas within radar coverage in the absence of spatial interpolation. 

The first region of interest is situated in northwest Australia, encompassing the Broome and Port Hedland radars. The high resolution climatology in Fig.\ \ref{maps}a reveals a pronounced land--sea contrast, with elevated hail frequencies beginning roughly 10--30 km inland from the coastline. This highlights the important role of sea breeze circulations for thunderstorm initiation in the tropical Australia \citep[\eg][]{Carbone00, May02}. The land--sea contrast may also reflect the climatologically high values of CIN offshore \citep{Riemann09} that acts to suppress convection even in the presence of strong low-level convergence \citep[\eg associated with land breezes;][]{Wapler12}. Fig.\ \ref{maps}f shows that once storms do initiate, most move toward the west, associated with easterly trade winds or monsoon return flow, depending on the time of year. While storms do initiate inland and generally move offshore, Fig.\ \ref{maps}a shows that almost all dissipate before reaching the coastline. Fig.\ \ref{stats}a also shows that hail streaks are, on average, shorter in northwest Australia, further substantiating the ephemeral nature of hailstorms in the region. The seasonal distribution of hailstorms in the region is distinctly bimodal (Fig.\ \ref{stats}b), with the break in December coinciding with the average arrival of the monsoon \citep{Hung04}. It is likely that increases in column-integrated moisture and weaker lapse rates associated with the active monsoon suppress hail activity during this period. It remains unclear whether MESH detections are substantiated by hailfall at the surface in these remote regions due to the extremely low population density. However, the coastal offset illustrated in Fig.\ \ref{maps}a may explain why tropical population centers on the coast rarely encounter hail, yet a surprising number of hail reports have been noted in the inland road network in northwest Australia \citep{Allen16}. 

The second region, located in northeastern Australia at a latitude similar to the first, provides an additional validation source for the characteristics of tropical hailstorms. Fig.\ \ref{maps}b shows that the overall hail frequency, and spatial offset from the coastline is similar to northwest Australia, albeit with increased offshore activity in the southern parts of the domain. Unlike coastal population centers on the west coast, non-zero hail frequencies extend to the coastline near Mackay, which notably recorded Australia's largest measured hailstone nearby in 2021 \citep[16 cm in diameter,][]{BoM20}. Fig.\ \ref{maps}g shows that storms predominantly move offshore toward the northeast, perhaps reflecting the southerly component required for significant instability aloft at these latitudes. These types of systems typically occur in late spring and early summer (Section \ref{s:national}), before the monsoon regime draws in tropical air masses from the east later in the season \citep{Sturman96}. Figs.\ \ref{stats}a and \ref{stats}c also confirm that, similar to the northwest Australia region, these tropical hail events are shorter in length and occur later in the day than their subtropical counterparts. The comparatively late diurnal cycle in the tropics (both northwestern and northeastern regions, usually between 4--6 pm local time) is consistent with that of overshooting convection in other locations in northern Australia such as Cairns and Darwin \citep{Bedka18b}.

The next region of interest coincides with the national hail frequency maximum in southeast Queensland/northeast New South Wales. This region contains the highest hail frequencies on the continent, consistent with prior hail climatologies \citep{Mcmaster01, Schuster05, Bedka18b}. The maximum hail frequency west of Casino in Fig.\ \ref{maps}c also aligns closely with the regional MESH-based climatology in \citet{Warren20}. It is difficult to compare the maximum frequency values observed (2.5 hail days per year per 100 km$^2$ in \citet{Warren20} versus $\sim$0.9 hailstorms per year here) due to methodological differences used for frequency estimation in both studies (see Section \ref{s:methods}\ref{norm}). The highest hail frequencies occur on either side of the QLD/NSW border, which follows the ridge top of the McPherson mountain range. The observed hail maxima along sloping terrain, rather than mountain peaks, is consistent with local thunderstorm initiation hypotheses, including thermally driven valley winds, orographic lifting, and leeward convergence of obstructed flows \citep{Sturman96, Soderholm17, Nisi18, Warren20}. A secondary hail maximum to the northwest of Brisbane was also observed by \citet{Soderholm17}, who hypothesized that storm intensification may be promoted by latent heat fluxes from nearby Lakes Wivenhoe and Somerset. Fig.\ \ref{maps}h indicates predominantly northeasterly storm motion, consistent with the dominant westerly steering flow in the subtropics and influence of southerly airmasses for high-end thunderstorm events in the region \citep{Soderholm19}. Previously identified southwest--northeast storm pathways may be identified in Fig.\ \ref{maps}c \citep[\eg from the border ranges toward Brisbane,][]{Soderholm17, Warren20}, along with a prominent storm pathway off the coast east of Casino. 

Hail frequencies are generally concentrated on the coastal plain to the east of the Great Dividing Range in Fig.\ \ref{maps}c. Along with the previously mentioned initiation mechanisms, this region is thought to be more favorable for hailstorms due to interactions with sea breezes (which supplement low-level moisture and wind shear), and more localized factors such as added latent heat fluxes from irrigated farmland and urban heat island effects \citep{Soderholm17, Soderholm17b}. Accordingly, hail frequencies reduce substantially onto the elevated New England Tableland in northeastern NSW, which is roughly traced by the 750 m elevation contour in Fig.\ \ref{maps}c.\footnote{Interestingly, older report-based climatologies indicate that the number of hail reports and crop losses are comparable between the New England Tableland and the coastal plain \citep{Mcmaster01, Schuster05}, likely a result of their earlier study periods.} Hail frequencies appear to drop spuriously in regions solely covered by the Grafton radar, which is known to display persistently poor calibration and hardware-related insensitivity (M. Curtis, 2023, personal communication, Bureau of Meteorology). Despite these issues, Grafton data were included in Fig.\ \ref{maps} to illustrate hail frequencies in a relative sense across northern NSW, due to the region's importance as a climatological hail maximum (see Fig.\ \ref{interp_freq}). The Grafton radar was not included in the national climatology (Appendix B) due to these reasons. 

The pattern of hail activity along the southern stretches of the NSW coastline, around Sydney, depicted in Fig.\ \ref{maps}d is generally similar to that in northeastern NSW. Hailstorms are concentrated on the coastal plain, with a particular prevalence on the eastern slopes of the Great Dividing Range. Hail frequencies appear to diminish atop elevated topography ($>$750 m), preferring mountain slopes rather than peaks. As a result, hail frequencies drop off considerably in the Snowy Mountains south of Canberra, and more broadly south of Sydney. \citet{Schuster05} also noted a marked drop in hail reports to Sydney's south, despite the large population density. Fig.\ \ref{maps}i supports the general picture of hail activity in the region, whereby storms initiate along the ranges to the east, before tracking eastward, intensifying across the coastal plain, and finally dissipating over the ocean \citep{Potts00, Schuster05, Warren20}. Oceanic hailstorms appear to be more common in the Sydney region compared to Brisbane \citep{Warren20}, aligning with the increased prevalence of lightning offshore in the region \citep{Dowdy14}.

The final region of interest is the Goldfields region in southwestern Australia (Fig.\ \ref{maps}e). The temperate climate and proximity to the southern ocean dictate that strong frontal systems are a dominant driver of storm activity in the region \citep{Sturman96, Allen14}. As a result of these broad-scale initiation mechanisms, hail frequencies are generally quite uniform spatially, with a less-pronounced land--sea contrast. The southeasterly directions shown in Fig.\ \ref{maps}j are consistent with the orientation of prefrontal troughs, which form ahead of southern ocean cold fronts roughly twice a month in summer \citep{Hanstrum90a}. These setups advect warm continental air toward the southeast during their passage across the southern parts of the continent, triggering severe thunderstorms under favorable environmental conditions \citep{Hanstrum90a, Hanstrum90b}, and are commonly responsible for hailstorms across southern Australia \citep{Buckley10}. Fig.\ \ref{stats}a. also shows the resulting hail streaks in southern WA are notably longer than those in the rest of the country.

\section{Summary and Future Directions}\label{s:conclusion}

This study aimed to develop a national radar-based hail climatology with a broad focus on the national- and regional-scales. Challenges unique to the Australian radar archive (\eg varying radar frequencies and limited spatial coverage) complicated these endeavors, and various methodological advances were required to achieve the objectives of the study. The main methodological contributions outlined in this study are as follows:

\begin{itemize}
    \item The introduction of an object-based hail processing algorithm enabled the calculation of resolution-invariant hail frequencies, which are more intuitive for individual observers and enable intercomparison between studies.
    \item A hail reflectivity correction was proposed for calculating Maximum Estimated Size of Hail (MESH) estimates at C-band, minimizing climatological differences with S-bands in the Australian archive.
    \item A novel variational method for interpolating radar-based hail frequencies across the country was introduced to produce a spatially continuous national hail climatology. 
\end{itemize}

These methodological advances permitted the investigation of hail occurrence across the aforementioned range of spatial scales. First, the hail frequencies were estimated on a national scale in Section \ref{s:national}, and showed a broad hail maximum on the east coast of the country between roughly Mackay and Sydney. The radar-based archive indicates that hailstorms occur once every 3--5 years at individual locations on the east coast. Elevated hail frequencies also occur on the northwest coast and inland stretches of southwest Australia, but are less than half as frequent as those on the east coast. 

Regional climatologies showed that hailstorms occur most frequently on the eastern slopes of the Great Dividing Range around the Brisbane and Sydney regions (one hail event every $\sim$1--2 years). Tropical hailstorms were found to be shorter in terms of length and initiate later in the day with a strong land--sea contrast. Subtropical hailstorms on the East coast were found to shift poleward throughout the summer, starting in Queensland in November and progressing toward Sydney by March. Finally, temperate hailstorms in southern Australia were observed to produce longer hail streaks and mainly propagate toward the southeast, corresponding with known synoptic-scale drivers in the region. 

During this study, several important avenues for future work were identified. First, secondary data sources, such as hail proxies from weather models or various satellite-based products, should be used to help constrain hail frequency estimates in areas with poor radar coverage, such as western New South Wales. These data sources may be incorporated as an additional background constraint in the variational interpolation method introduced herein. Further research should also investigate the accuracy of MESH for hail identification in different climatic regions such as the tropics and southern Australia, using hail both hail reports and comparisons with satellite-based remote sensing products. These studies should also investigate how storm-driven hail advection, melting, and size sorting can also affect radar-based climatological assessments \citep[\eg][]{Brook21}. Finally, future research should also explore how the radar archive may be used to identify long-term trends and interannual/decadal variability previously noted in the Australian hail climatology \citep{Schuster05, Allen14}, in the context of how they may be evolving in a changing climate.

\begin{figure*}
    \begin{minipage}{\textwidth}
    \appendix[B] 
    \appendixtitle{Radar Information}
    
    \captionof{table}{Relevant information for radars used in this study. Missing end dates indicate radars were still operational at the end of the study period, altitudes indicate antenna height above sea level, and beam widths are measured in degrees to half-power. Refer to \texttt{https://openradar.io/} for further details.} \label{table1}

    \footnotesize
    \setlength\extrarowheight{-1pt}
    \BeforeBeginEnvironment{tabular}{\footnotesize}%
        
    \begin{tabularx}{\textwidth}{llcccccccc}
	\toprule
           Location &    Radar Type & Start Date &   End Date &  Dur. (yr) &   Lat. ($^{\circ}$S)&   Lon. ($^{\circ}$E)&  Alt. (m) & Freq. &  BWidth ($^{\circ}$) \\
\midrule
 Adelaide (B. Park) & Meteor1500SDP & 27/10/2005 &          - &       16.2 & -34.62 & 138.47 &      29.0 &     S &                  1.0 \\
 Adelaide (S. Hill) &        WSR81C & 01/01/2000 &          - &       22.0 & -35.33 & 138.50 &     395.1 &     C &                  1.7 \\
             Albany &  Meteor735CDP & 21/02/2003 &          - &       18.9 & -34.94 & 117.82 &      93.0 &     C &                  1.0 \\
      Alice Springs &      WF100-6C & 28/10/2003 &          - &       18.2 & -23.80 & 133.89 &     563.3 &     C &                  1.7 \\
         Bairnsdale & Wurrung-2502C & 06/05/2008 &          - &       13.7 & -37.89 & 147.58 &      65.3 &     C &                  1.7 \\
              Bowen &      WF100-6C & 07/05/2004 &          - &       17.7 & -19.89 & 148.07 &      72.5 &     C &                  1.7 \\
 Brisbane (Marburg) &        WSR74S & 01/01/2000 &          - &       22.0 & -27.61 & 152.54 &     371.1 &     S &                  1.9 \\
Brisbane (Mt Stap.) & Meteor1500SDP & 08/06/2006 &          - &       15.6 & -27.72 & 153.24 &     175.0 &     S &                  1.0 \\
             Broome &     DWSR2502C & 18/03/2008 &          - &       13.8 & -17.95 & 122.24 &      31.4 &     C &                  1.7 \\
             Cairns &     DWSR2502C & 11/01/2000 &          - &       22.0 & -16.82 & 145.68 &     660.8 &     C &                  1.0 \\
           Canberra &       DWSR74S & 22/11/2002 &          - &       19.1 & -35.66 & 149.51 &    1382.7 &     S &                  1.9 \\
          Carnarvon &     DWSR2502C & 07/12/2004 &          - &       17.1 & -24.89 & 113.67 &      12.0 &     C &                  1.7 \\
             Ceduna &     DWSR2502C & 06/01/2002 &          - &       20.0 & -32.13 & 133.70 &      32.0 &     C &                  1.7 \\
            Dampier &     WRM200CDP & 13/08/2009 &          - &       12.4 & -20.65 & 116.68 &      55.0 &     C &                  1.0 \\
  Darwin (Berrimah) &     DWSR2502C & 07/12/2000 &          - &       21.1 & -12.46 & 130.93 &      60.0 &     C &                  1.0 \\
            Emerald &     DWSR8502S & 09/03/2010 &          - &       11.8 & -23.55 & 148.24 &     211.3 &     S &                  1.9 \\
          Esperence &  Meteor735CDP & 19/02/2003 &          - &       18.9 & -33.83 & 121.89 &      42.9 &     C &                  1.0 \\
          Geraldton &  Meteor735CDP & 08/02/2003 &          - &       18.9 & -28.80 & 114.70 &      47.7 &     C &                  1.0 \\
              Giles &      WF100-5C & 15/10/2003 &          - &       18.2 & -25.03 & 128.30 &     607.4 &     C &                  2.0 \\
          Gladstone &        WSR74S & 01/01/2000 &          - &       22.0 & -23.86 & 151.26 &      92.5 &     S &                  1.9 \\
               Gove &      TVDR2500 & 01/10/2002 &          - &       19.3 & -12.28 & 136.82 &      75.7 &     C &                  1.7 \\
            Grafton &        WSR74S & 01/01/2000 &          - &       22.0 & -29.62 & 152.95 &      49.8 &     S &                  1.9 \\
             Gympie &     DWSR8502S & 01/01/2000 &          - &       22.0 & -25.96 & 152.58 &     355.0 &     S &                  1.9 \\
        Halls Creek &      WF100-5C & 26/03/2008 &          - &       13.8 & -18.23 & 127.66 &     439.4 &     C &                  1.7 \\
             Hobart &     DWSR2502C & 22/03/2012 &          - &        9.8 & -43.11 & 147.81 &     511.7 &     C &                  1.0 \\
         Kalgoorlie &     DWSR2502C & 13/11/2003 &          - &       18.1 & -30.78 & 121.45 &     390.0 &     C &                  1.0 \\
 Katherine (Tindal) &        WSR81C & 15/10/2003 &          - &       18.2 & -14.51 & 132.45 &     148.4 &     C &                  1.7 \\
          Learmouth &      TVDR2500 & 11/02/2008 &          - &       13.9 & -22.10 & 114.00 &     333.3 &     C &                  1.7 \\
          Longreach &      WF100-6C & 18/12/2000 &          - &       21.0 & -23.43 & 144.29 &     201.6 &     C &                  1.7 \\
             Mackay &      TVDR2500 & 01/01/2000 &          - &       22.0 & -21.12 & 149.22 &      43.9 &     C &                  1.7 \\
          Melbourne & Meteor1500SDP & 01/01/2000 &          - &       22.0 & -37.86 & 144.76 &      45.0 &     S &                  1.0 \\
            Mildura &      WF100-5C & 01/01/2002 & 15/03/2021 &       19.2 & -34.24 & 142.09 &      59.3 &     C &                  2.0 \\
              Moree &      WF100-5C & 01/01/2000 &          - &       22.0 & -29.49 & 149.85 &     221.5 &     C &                  1.7 \\
  Mornington Island &        WSR81C & 01/01/2000 &          - &       22.0 & -16.67 & 139.17 &      19.6 &     C &                  1.7 \\
          Mount Isa &     DWSR8502S & 14/09/2012 &          - &        9.3 & -20.71 & 139.56 &     516.7 &     S &                  1.9 \\
      N.W. Tasmania & Wurrung-2502C & 01/01/2000 &          - &       22.0 & -41.18 & 145.58 &     601.0 &     C &                  1.0 \\
              Namoi &     DWSR8502S & 02/06/2010 &          - &       11.6 & -31.02 & 150.19 &     720.0 &     S &                  1.9 \\
          Newcastle &       DWSR74S & 01/01/2000 &          - &       22.0 & -32.73 & 152.03 &      84.0 &     S &                  1.9 \\
 Perth (Serpentine) &      TVDR2500 & 02/02/2010 &          - &       11.9 & -32.39 & 115.87 &      39.8 &     C &                  1.0 \\
       Port Hedland &      TVDR2500 & 12/03/2008 &          - &       13.8 & -20.37 & 118.63 &      23.8 &     C &                  1.7 \\
        Rockhampton &        WF100C & 01/01/2000 & 31/08/2014 &       14.7 & -23.38 & 150.47 &      14.0 &     C &                  1.6 \\
   South Doodlakine & Wurrung-2502C & 15/02/2017 &          - &        4.9 & -31.78 & 117.95 &     416.0 &     C &                  1.7 \\
   Sydney (Kurnell) &      WF100-6C & 01/01/2000 &          - &       22.0 & -34.01 & 151.23 &      63.8 &     C &                  1.0 \\
  Sydney (T. Hills) & Meteor1500SDP & 15/05/2009 &          - &       12.6 & -33.70 & 151.21 &     223.5 &     S &                  1.0 \\
      Tennant Creek &        WF100C & 10/07/2007 & 30/06/2015 &        8.0 & -19.64 & 134.18 &     394.0 &     C &                  1.8 \\
         Townsville &     DWSR2502C & 03/08/2011 &          - &       10.4 & -19.42 & 146.55 &     612.5 &     C &                  1.0 \\
            Warrego &      TVDR2500 & 06/10/2006 &          - &       15.2 & -26.44 & 147.35 &     534.4 &     C &                  1.7 \\
            Warruwi &     DWSR2502C & 15/11/2012 &          - &        9.1 & -11.65 & 133.38 &      43.7 &     C &                  1.0 \\
              Weipa &        WF100C & 01/01/2000 & 11/10/2015 &       15.8 & -12.67 & 141.92 &      35.0 &     C &                  1.7 \\
         Wollongong &     DWSR8502S & 01/01/2000 &          - &       22.0 & -34.26 & 150.88 &     471.6 &     S &                  1.9 \\
            Woomera &      WF100-6C & 04/01/2001 &          - &       21.0 & -31.16 & 136.80 &     175.9 &     C &                  1.7 \\
         Yarrawonga &      WF100-6C & 25/06/2002 &          - &       19.5 & -36.03 & 146.02 &     146.0 &     C &                  1.0 \\
\bottomrule

\end{tabularx}        
\end{minipage}
\end{figure*}

\appendix[A] 

\appendixtitle{Iterative Orthogonal Linear Regression}

We expect the reflectivity response of large hail to diminish at C-band due the theoretical considerations outlined in Section \ref{s:freq}. Here, we attempt to correct C-band hail reflectivities to match that at S-band, where the chosen hail detection methodology was developed. The fitting procedure used to generate the correction is a linear orthogonal regression. This technique is suitable for our purposes, as it considers measurement error in both $x$ and $y$ variables, which is distinct from an ordinary least squares regression that assumes the $x$ variable is known without error. Orthogonal regression is also suitable for our application due to our use of point-matched values of the same variable, which, due to the identical matching procedure employed to both, we may assume to have equal variance \citep{York66}. \citet{Carrol96} show that orthogonal regression may be misused if the ratio of input variable variance is incorrectly specified, or due to nonlinearities in the underlying dataset. Despite these concerns, we consider it an appropriate technique for our case and employ an efficient FORTRAN implementation \citep{Boggs89}.

The aim is to fit only those reflectivities that correspond to MESH detections, which dictates data is taken from above the 0$^{\circ}$C isotherm in columns where MESH exceeds 10 mm (refer to Section \ref{s:freq} for a discussion on the 10-mm MESH threshold). We also only use data above 40 dBZ, as in the MESH formulation. Orthogonal regression is used to fit C- and S-band reflectivities as the $x$ and $y$ variables, respectively, resulting in a standard linear function : $y = mx+c$, where $m$ and $c$ are the linear coefficients. When the correction is applied to the $y$ data, the initial fit is transposed as follows:

\begin{equation}\label{fit1}
    Y_1 = \frac{y_0 - c_1}{m_1}
\end{equation}

\noindent where $Y$ denotes the corrected $y$ data, $y_0$ is the initial $y$ data and subscripts denote the n$^{\text{th}}$ correction. Depending on the sign of the correction, $Y_1$ may now contain data above or below the 40 dBZ threshold, which should be included or excluded from the fitting calculation, respectively. This problem is analogous to one encountered in radar calibration, where spaceborne reflectivity data above a certain threshold are used to calculate a calibration offset \citep{Protat11, Warren18}. The solution arises by proceeding iteratively, applying the correction and filtering the data according to the threshold at each step. The linear corrections in our case introduce additional complexity (compared to the constant offset used in radar calibration), and we now describe how the final correction may be obtained from the fit at each iteration. Once the 40 dBZ threshold has been applied to $Y_1$, we recorrect the data in the second iteration:

\begin{equation}\label{fit2}
    Y_2 = \frac{Y_1 - c_2}{m_2} = \frac{\frac{y_0 - c_1}{m_1} - c_2}{m_2}
\end{equation}

\noindent where Equation \ref{fit1} was substituted in the final expression. The refitting process continues until the difference between the coefficients $m$ and $c$ in consecutive iterations is less than a user-defined convergence criterion ($\epsilon$, where $\epsilon =1\times10^{-4}$ in this study). The resulting correction when applying the refitting procedure for $n$ iterations is generalized below,

\begin{equation} \label{fitn}
    Y_n = \frac{y_o}{\prod_{i=1}^nm_i} - \sum_{i=1}^{n}\frac{c_i}{\prod_{j=i}^nm_j}.
\end{equation}

The optimization converged in roughly 20 iterations for the fits illustrated in Fig.\ \ref{freq_pic}. Although no divergent cases were observed in our analysis, we offer no formal proof of convergence for the method outlined here.


\acknowledgments
This research was undertaken with the assistance of resources and services from the National Computational Infrastructure (NCI), which is supported by the Australian Government. Author Brook would like to acknowledge Guy Carpenter \& Company, LLC for industry funding throughout their postgraduate studies. We also acknowledge Mark Curtis and  Valentin Louf for discussions on the Australian Radar Archive.

%
%
\datastatement
Download instructions for all data used in this study may be found in the Australian Unified Radar Archive \citep[AURA,][]{Level1b} and are available at \url{https://www.openradar.io/operational-network}. Data processing software is available upon request to the author.


\bibliographystyle{ametsocV6}
\bibliography{newfile}

\begin{thebibliography}{112}
\providecommand{\natexlab}[1]{#1}
\providecommand{\url}[1]{\texttt{#1}}
\renewcommand{\UrlFont}{\rmfamily}
\providecommand{\urlprefix}{URL }
\expandafter\ifx\csname urlstyle\endcsname\relax
  \providecommand{\doi}[1]{https://doi.org/\discretionary{}{}{}#1}\else
  \providecommand{\doi}{https://doi.org/\discretionary{}{}{}\begingroup
  \urlstyle{rm}\Url}\fi
\providecommand{\eprint}[2][]{\url{#2}}

\bibitem[{Allen(2012)}]{Allen12}
Allen, J., 2012: {Supercell storms: Melbourne's white Christmas 2011.}
  \textit{Bull. Aust. Meteorol. Oceanogr. Soc.}, \textbf{25}, 47--51.

\bibitem[{Allen and Allen(2016)Allen, and Allen}]{Allen16}
Allen, J.~T., and E.~R. Allen, 2016: {A review of severe thunderstorms in
  Australia}. \textit{Atmos. Res.}, \textbf{178-179~(Supplement C)}, 347--366,
  \doi{https://doi.org/10.1016/j.atmosres.2016.03.011}.

\bibitem[{Allen and Karoly(2014)Allen, and Karoly}]{Allen14}
Allen, J.~T., and D.~J. Karoly, 2014: {A climatology of Australian severe
  thunderstorm environments 1979-2011: Inter-annual variability and ENSO
  influence}. \textit{Int. J. Climatol.}, \textbf{34~(1)}, 81--97,
  \doi{10.1002/joc.3667}.

\bibitem[{Allen et~al.(2015)Allen, Tippett, Ashley, Kumjian,, and
  Cavanaugh}]{Allen15}
Allen, J.~T., M.~K. Tippett, W.~S. Ashley, M.~R. Kumjian, and D.~Cavanaugh,
  2015: {The characteristics of United States hail reports: 1955-2014}.
  \textit{Electron. J. Sev. Storms Meteor.}, \textbf{10~(3)}, 1--31.

\bibitem[{Anagnostou and Krajewski(1999)Anagnostou, and
  Krajewski}]{Anagnostou99}
Anagnostou, E.~N., and W.~F. Krajewski, 1999: {Real-time radar rainfall
  estimation. Part I: Algorithm formulation}. \textit{J. Atmos. Ocean.
  Technol.}, \textbf{16~(2)}, 189--197,
  \doi{10.1175/1520-0426(1999)016<0189:RTRREP>2.0.CO;2}.

\bibitem[{{Australian Bureau of Statistics}(2021)}]{ABS21}
{Australian Bureau of Statistics}, 2021: {Regional population}. {Accessed 8
  April 2023,
  \url{https://www.abs.gov.au/statistics/people/population/regional-population/latest-release}}.

\bibitem[{Bang and Cecil(2019)Bang, and Cecil}]{Bang19}
Bang, S.~D., and D.~J. Cecil, 2019: {Constructing a multifrequency passive
  microwave hail retrieval and climatology in the GPM domain}. \textit{J. Appl.
  Meteor. Climatol.}, \textbf{58~(9)}, 1889--1904,
  \doi{10.1175/JAMC-D-19-0042.1}.

\bibitem[{Bang and Cecil(2021)Bang, and Cecil}]{Bang21}
Bang, S.~D., and D.~J. Cecil, 2021: {Testing passive microwave-based hail
  retrievals using GPM DPR Ku-band radar}. \textit{J. Appl. Meteor. Climatol.},
  \textbf{60~(3)}, 255--271, \doi{https://doi.org/10.1175/JAMC-D-20-0129.1}.

\bibitem[{Barth et~al.(2014)Barth, Beckers, Troupin, Alvera-Azc{\'{a}}rate,,
  and Vandenbulcke}]{Barth14}
Barth, A., J.-M. Beckers, C.~Troupin, A.~Alvera-Azc{\'{a}}rate, and
  L.~Vandenbulcke, 2014: {Divand-1.0: n-dimensional variational data analysis
  for ocean observations}. \textit{Geosci. Model Dev.}, \textbf{7~(1)},
  225--241, \doi{10.5194/gmd-7-225-2014}.

\bibitem[{Battan(1971)}]{Battan71}
Battan, L.~J., 1971: {Radar attenuation by wet ice spheres}. \textit{J. Appl.
  Meteor. Climatol.}, \textbf{10~(2)}, 247--252,
  \doi{https://doi.org/10.1175/1520-0450(1971)010<0247:RABWIS>2.0.CO;2}.

\bibitem[{Beckers et~al.(2014)Beckers, Barth, Troupin,, and
  Alvera-Azc{\'{a}}rate}]{Beckers14}
Beckers, J.-M., A.~Barth, C.~Troupin, and A.~Alvera-Azc{\'{a}}rate, 2014:
  {Approximate and efficient methods to assess error fields in spatial gridding
  with data interpolating variational analysis (DIVA)}. \textit{J. Atmos.
  Ocean. Technol.}, \textbf{31~(2)}, 515--530,
  \doi{https://doi.org/10.1175/JTECH-D-13-00130.1}.

\bibitem[{Bedka et~al.(2010)Bedka, Brunner, Dworak, Feltz, Otkin,, and
  Greenwald}]{Bedka10}
Bedka, K., J.~Brunner, R.~Dworak, W.~Feltz, J.~Otkin, and T.~Greenwald, 2010:
  {Objective satellite-based detection of overshooting tops using infrared
  window channel brightness temperature gradients}. \textit{J. Appl. Meteor.
  Climatol.}, \textbf{49~(2)}, 181--202, \doi{10.1175/2009JAMC2286.1}.

\bibitem[{Bedka(2011)}]{Bedka11}
Bedka, K.~M., 2011: {Overshooting cloud top detections using MSG SEVIRI
  infrared brightness temperatures and their relationship to severe weather
  over Europe}. \textit{Atmos. Res.}, \textbf{99~(2)}, 175--189,
  \doi{https://doi.org/10.1016/j.atmosres.2010.10.001}.

\bibitem[{Bedka et~al.(2018)Bedka, Allen, Punge, Kunz,, and
  Simanovic}]{Bedka18b}
Bedka, K.~M., J.~T. Allen, H.~J. Punge, M.~Kunz, and D.~Simanovic, 2018: {A
  long-term overshooting convective cloud top detection database over Australia
  derived from MTSAT Japanese advanced meteorological Imager observations}.
  \textit{J. Appl. Meteor. Climatol.}, \textbf{57~(0)}, 937--951,
  \doi{10.1175/JAMC-D-17-0056.1}.

\bibitem[{Bednarczyk and Sousounis(2012)Bednarczyk, and
  Sousounis}]{Bednarczyk12}
Bednarczyk, C., and P.~Sousounis, 2012: {Hail climatology of Australia based on
  lightning and reanalysis}. \textit{27th Conference on Severe Local Storms},
  Amer. Meteor. Soc.,
  \urlprefix\url{https://ams.confex.com/ams/27SLS/webprogram/Manuscript/Paper255889/extendedAbstract_pdf.pdf}.

\bibitem[{Boggs et~al.(1989)Boggs, Donaldson, Byrd,, and Schnabel}]{Boggs89}
Boggs, P.~T., J.~R. Donaldson, R.~h. Byrd, and R.~B. Schnabel, 1989: {Algorithm
  676: ODRPACK: software for weighted orthogonal distance regression}.
  \textit{ACM Trans. Math. Softw.}, \textbf{15~(4)}, 348--364,
  \doi{10.1145/76909.76913}.

\bibitem[{{BoM}(2021)}]{BoM20}
{BoM}, 2021: {Bureau of meteorology severe thunderstorm archive}. Aust. Bureau
  of Meteorology, {Accessed 21 June 2023,
  \url{http://www.bom.gov.au/australia/stormarchive/}}.

\bibitem[{Borowska et~al.(2011)Borowska, Ryzhkov, Zrni{\'{c}}, Simmer,, and
  Palmer}]{Borowska11}
Borowska, L., A.~Ryzhkov, D.~Zrni{\'{c}}, C.~Simmer, and R.~Palmer, 2011:
  {Attenuation and differential attenuation of 5-cm wavelength radiation in
  melting hail}. \textit{J. Appl. Meteor. Climatol.}, \textbf{50~(1)}, 59--76,
  \doi{https://doi.org/10.1175/2010JAMC2465.1}.

\bibitem[{Briggs(1974)}]{Briggs74}
Briggs, I.~C., 1974: {Machine contouring using minimum curvature}.
  \textit{Geophysics}, \textbf{39~(1)}, 39--48, \doi{10.1190/1.1440410}.

\bibitem[{Brook et~al.(2023)Brook, Protat, Potvin, Soderholm,, and
  McGowan}]{Brook23}
Brook, J.~P., A.~Protat, C.~K. Potvin, J.~S. Soderholm, and H.~McGowan, 2023:
  {The effects of spatial interpolation on a novel, dual-Doppler 3D wind
  retrieval technique}, \doi{10.48550/ARXIV.2301.07913}.

\bibitem[{Brook et~al.(2021)Brook, Protat, Soderholm, Carlin, McGowan,, and
  Warren}]{Brook21}
Brook, J.~P., A.~Protat, J.~Soderholm, J.~T. Carlin, H.~McGowan, and R.~A.
  Warren, 2021: {Hailtrack-improving radar-based hailfall estimates by modeling
  hail trajectories}. \textit{J. Appl. Meteor. Climatol.}, \textbf{60~(3)},
  237--254.

\bibitem[{Brook et~al.(2022)Brook, Protat, Soderholm, Warren,, and
  McGowan}]{Brook22}
Brook, J.~P., A.~Protat, J.~S. Soderholm, R.~A. Warren, and H.~McGowan, 2022:
  {A variational interpolation method for gridding weather radar data}.
  \textit{J. Atmos. Ocean. Technol.}, \textbf{39~(11)}, 1633–1654,
  \doi{10.1175/JTECH-D-22-0015.1}.

\bibitem[{Buckley et~al.(2010)Buckley, Sullivan, Chan,, and
  Leplastrier}]{Buckley10}
Buckley, B.~W., W.~Sullivan, P.~Chan, and M.~Leplastrier, 2010: {Two record
  breaking Australian hail storms: storm environments, damage characteristics
  and rarity}. \textit{25th Conference on Severe Local Storms}, American
  Meteorological Society Denver, Colorado, 11--14.

\bibitem[{Bunkers and Smith(2013)Bunkers, and Smith}]{Bunkers13}
Bunkers, M.~J., and P.~L. Smith, 2013: {Comments on “An objective
  high-resolution hail climatology of the contiguous united States”}.
  \textit{Wea. Forecasting}, \textbf{28~(3)}, 915--917,
  \doi{https://doi.org/10.1175/WAF-D-13-00020.1}.

\bibitem[{Cao et~al.(2013)Cao, Hong, Qi, Wen, Zhang, Gourley,, and
  Liao}]{Cao13}
Cao, Q., Y.~Hong, Y.~Qi, Y.~Wen, J.~Zhang, J.~J. Gourley, and L.~Liao, 2013:
  {Empirical conversion of the vertical profile of reflectivity from Ku-band to
  S-band frequency}. \textit{J. Geophys. Res. Atmos.}, \textbf{118~(4)},
  1814--1825, \doi{https://doi.org/10.1002/jgrd.50138}.

\bibitem[{Carbone et~al.(2000)Carbone, Wilson, Keenan,, and Hacker}]{Carbone00}
Carbone, R.~E., J.~W. Wilson, T.~D. Keenan, and J.~M. Hacker, 2000: {Tropical
  island convection in the absence of significant topography. Part I: Life
  cycle of diurnally forced convection}. \textit{Mon. Wea. Rev.},
  \textbf{128~(10)}, 3459--3480,
  \doi{https://doi.org/10.1175/1520-0493(2000)128<3459:TICITA>2.0.CO;2}.

\bibitem[{Carroll and Ruppert(1996)Carroll, and Ruppert}]{Carrol96}
Carroll, R.~J., and D.~Ruppert, 1996: {The use and misuse of orthogonal
  regression in linear errors-in variables models}. \textit{Amer. Stat.},
  \textbf{50~(1)}, 1--6, \doi{10.2307/2685035}.

\bibitem[{Cecchini et~al.(2022)Cecchini, Heymsfield, Honeyager, Field,
  Machado,, and Dias}]{Cecchini22}
Cecchini, M.~A., A.~J. Heymsfield, R.~Honeyager, P.~Field, L.~A.~T. Machado,
  and M.~A. F. d.~S. Dias, 2022: {Revisiting the hail radar
  reflectivity–kinetic energy flux relation by combining t-matrix and
  discrete dipole approximation calculations to size distribution
  observations}. \textit{J. Atmos. Sci.}, \textbf{79~(7)}, 1927--1940,
  \doi{https://doi.org/10.1175/JAS-D-20-0373.1}.

\bibitem[{Cecil(2009)}]{Cecil09}
Cecil, D.~J., 2009: {Passive microwave brightness temperatures as proxies for
  hailstorms}. \textit{J. Appl. Meteor. Climatol.}, \textbf{48~(6)},
  1281--1286, \doi{10.1175/2009JAMC2125.1}.

\bibitem[{Cecil and Blankenship(2012)Cecil, and Blankenship}]{Cecil12}
Cecil, D.~J., and C.~B. Blankenship, 2012: {Toward a global climatology of
  severe hailstorms as estimated by satellite passive microwave imagers}.
  \textit{J. Clim.}, \textbf{25~(2)}, 687--703,
  \doi{10.1175/JCLI-D-11-00130.1}.

\bibitem[{Changnon(1970)}]{Changon70}
Changnon, S.~A., 1970: {Hailstreaks}. \textit{J. Atmos. Sci.}, \textbf{27~(1)},
  109--125, \doi{10.1175/1520-0469(1970)027<0109:H>2.0.CO;2}.

\bibitem[{Changnon et~al.(2009)Changnon, Changnon,, and Hilberg}]{Changnon09}
Changnon, S.~A., D.~Changnon, and S.~D. Hilberg, 2009: {Hailstorms across the
  nation: an atlas about hail and its damages}. Illinois State Water Survey
  Contract Report CR-2009-12.

\bibitem[{Cintineo et~al.(2012)Cintineo, Smith, Lakshmanan, Brooks,, and
  Ortega}]{Cintineo12}
Cintineo, J.~L., T.~M. Smith, V.~Lakshmanan, H.~E. Brooks, and K.~L. Ortega,
  2012: {An objective high-resolution hail climatology of the contiguous united
  states}. \textit{Wea. Forecasting}, \textbf{27~(5)}, 1235--1248,
  \doi{10.1175/WAF-D-11-00151.1}.

\bibitem[{Cotton et~al.(2010)Cotton, Bryan,, and van~den Heever}]{Cotton10}
Cotton, W.~R., G.~Bryan, and S.~C. van~den Heever, 2010: \textit{{Storm and
  cloud dynamics}}. 2nd ed., Elsevier Science, Burlington.

\bibitem[{Dahl et~al.(2019)Dahl, Shapiro, Potvin, Theisen, Gebauer, Schenkman,,
  and Xue}]{Dahl19}
Dahl, N.~A., A.~Shapiro, C.~K. Potvin, A.~Theisen, J.~G. Gebauer, A.~D.
  Schenkman, and M.~Xue, 2019: {High-resolution, rapid-scan dual-Doppler
  retrievals of vertical velocity in a simulated supercell}. \textit{J. Atmos.
  Ocean. Technol.}, \textbf{36~(8)}, 1477--1500,
  \doi{10.1175/JTECH-D-18-0211.1}.

\bibitem[{Dowdy and Kuleshov(2014)Dowdy, and Kuleshov}]{Dowdy14}
Dowdy, A., and Y.~Kuleshov, 2014: {Lightning climatology of Australia: temporal
  and spatial variability}. \textit{Aust. Meteorol. Oceanogr. J.}, \textbf{64},
  9--14.

\bibitem[{Dowdy et~al.(2020)Dowdy, Soderholm, Brook, Brown,, and
  McGowan}]{Dowdy20}
Dowdy, A.~J., J.~Soderholm, J.~Brook, A.~Brown, and H.~McGowan, 2020:
  {Quantifying hail and lightning risk factors using long-term observations
  around Australia}. \textit{J. Geophys. Res. Atmos.}, \textbf{125~(21)},
  2020JD033\,101, \doi{https://doi.org/10.1029/2020JD033101}.

\bibitem[{Drosdowsky and Holland(1987)Drosdowsky, and Holland}]{Drosdowsky87}
Drosdowsky, W., and G.~J. Holland, 1987: {North Australian cloud lines}.
  \textit{Mon. Wea. Rev.}, \textbf{115~(11)}, 2645--2659.

\bibitem[{Dworak et~al.(2012)Dworak, Bedka, Brunner,, and Feltz}]{Dworak12}
Dworak, R., K.~Bedka, J.~Brunner, and W.~Feltz, 2012: {Comparison between
  goes-12 overshooting-top detections, WSR-88D radar reflectivity, and severe
  storm reports}. \textit{Wea. Forecasting}, \textbf{27~(3)}, 684--699,
  \doi{10.1175/WAF-D-11-00070.1}.

\bibitem[{Eichner(2019)}]{Munich19}
Eichner, J., 2019: {Icy cricket balls from above}. {Accessed 7 April 2023,
  \url{https://www.munichre.com/topics-online/en/climate-change-and-natural-disasters/climate-change/Icy-cricket-balls-from-above.html}}.

\bibitem[{Fabry et~al.(2017)Fabry, Meunier, Treserras, Cournoyer,, and
  Nelson}]{Fabry17}
Fabry, F., V.~Meunier, B.~P. Treserras, A.~Cournoyer, and B.~Nelson, 2017: {On
  the climatological use of radar data mosaics: possibilities and challenges}.
  \textit{Bull. Amer. Meteor. Soc.}, \textbf{98~(10)}, 2135--2148,
  \doi{https://doi.org/10.1175/BAMS-D-15-00256.1}.

\bibitem[{Ferraro et~al.(2015)Ferraro, Beauchamp, Cecil,, and
  Heymsfield}]{Ferraro15}
Ferraro, R., J.~Beauchamp, D.~Cecil, and G.~Heymsfield, 2015: {A prototype hail
  detection algorithm and hail climatology developed with the advanced
  microwave sounding unit (AMSU)}. \textit{Atmos. Res.}, \textbf{163}, 24--35.

\bibitem[{Foote(1984)}]{Foote84}
Foote, G.~B., 1984: {A study of hail growth utilizing observed storm
  conditions}. \textit{J. Climate Appl. Meteor.}, \textbf{23~(1)}, 84--101,
  \doi{10.1175/1520-0450(1984)023<0084:ASOHGU>2.0.CO;2}.

\bibitem[{Frisby and Sansom(1967)Frisby, and Sansom}]{Frisby67}
Frisby, E.~M., and H.~W. Sansom, 1967: {Hail incidence in the tropics}.
  \textit{J. Appl. Meteor.}, \textbf{6~(2)}, 339--354,
  \doi{10.1175/1520-0450(1967)006<0339:HIITT>2.0.CO;2}.

\bibitem[{Gabella and Notarpietro(2002)Gabella, and Notarpietro}]{Gabella02}
Gabella, M., and R.~Notarpietro, 2002: {Ground clutter characterization and
  elimination in mountainous terrain}. \textit{Proceedings of ERAD}, 305--311,
  \url{https://www.copernicus.org/erad/online/erad-305.pdf}.

\bibitem[{Goldstein and Osher(2009)Goldstein, and Osher}]{Goldstein09}
Goldstein, T., and S.~Osher, 2009: {The split Bregman method for L1-regularized
  problems}. \textit{SIAM J. Imaging Sci.}, \textbf{2~(2)}, 323--343,
  \doi{10.1137/080725891}.

\bibitem[{Gu et~al.(2011)Gu, Ryzhkov, Zhang, Neilley, Knight, Wolf,, and
  Lee}]{Gu11}
Gu, J.-Y., A.~Ryzhkov, P.~Zhang, P.~Neilley, M.~Knight, B.~Wolf, and D.-I. Lee,
  2011: {Polarimetric attenuation correction in heavy rain at C band}.
  \textit{J. Appl. Meteor. Climatol.}, \textbf{50~(1)}, 39--58,
  \doi{https://doi.org/10.1175/2010JAMC2258.1}.

\bibitem[{Gunturi and Tippett(2017)Gunturi, and Tippett}]{Gunturi17}
Gunturi, P., and M.~K. Tippett, 2017: {Managing severe thunderstorm risk:
  Impact of ENSO on US tornado and hail frequencies}. Tech. rep., Tech. Rep.

\bibitem[{Hanstrum et~al.(1990{\natexlab{a}})Hanstrum, Wilson,, and
  Barrell}]{Hanstrum90a}
Hanstrum, B.~N., K.~J. Wilson, and S.~L. Barrell, 1990{\natexlab{a}}:
  {Prefrontal troughs over southern Australia. Part I: a climatology}.
  \textit{Wea. Forecasting}, \textbf{5~(1)}, 22--31,
  \doi{https://doi.org/10.1175/1520-0434(1990)005<0022:PTOSAP>2.0.CO;2}.

\bibitem[{Hanstrum et~al.(1990{\natexlab{b}})Hanstrum, Wilson,, and
  Barrell}]{Hanstrum90b}
Hanstrum, B.~N., K.~J. Wilson, and S.~L. Barrell, 1990{\natexlab{b}}:
  {Prefrontal troughs over southern Australia. Part II: a case study of
  frontogenesis}. \textit{Wea. Forecasting}, \textbf{5~(1)}, 32--46,
  \doi{https://doi.org/10.1175/1520-0434(1990)005<0032:PTOSAP>2.0.CO;2}.

\bibitem[{Heistermann et~al.(2013)Heistermann, Jacobi,, and
  Pfaff}]{Heistermann13}
Heistermann, M., S.~Jacobi, and T.~Pfaff, 2013: {Technical note: an open source
  library for processing weather radar data (wradlib)}. \textit{Hydrol. Earth
  Syst. Sci.}, \textbf{17~(2)}, 863--871, \doi{10.5194/hess-17-863-2013}.

\bibitem[{Homeyer and Bowman(2017)Homeyer, and Bowman}]{Homeyer17b}
Homeyer, C.~R., and K.~P. Bowman, 2017: {Algorithm description document for
  version 3.1 of the three-dimensional gridded NEXRAD WSR-88D radar (GridRad)
  dataset}. Tech. rep.

\bibitem[{Homeyer et~al.(2017)Homeyer, McAuliffe,, and Bedka}]{Homeyer17}
Homeyer, C.~R., J.~D. McAuliffe, and K.~M. Bedka, 2017: {On the development of
  above-anvil cirrus plumes in extratropical convection}. \textit{J. Atmos.
  Sci.}, \textbf{74~(5)}, 1617--1633, \doi{10.1175/JAS-D-16-0269.1}.

\bibitem[{Hung and Yanai(2004)Hung, and Yanai}]{Hung04}
Hung, C.-W., and M.~Yanai, 2004: {Factors contributing to the onset of the
  Australian summer monsoon}. \textit{Q. J. R. Meteor. Soc.},
  \textbf{130~(597)}, 739--758, \doi{https://doi.org/10.1256/qj.02.191}.

\bibitem[{Kaltenboeck and Ryzhkov(2013)Kaltenboeck, and
  Ryzhkov}]{Kaltenboeck13}
Kaltenboeck, R., and A.~Ryzhkov, 2013: {Comparison of polarimetric signatures
  of hail at S and C bands for different hail sizes}. \textit{Atmos. Res.},
  \textbf{123}, 323--336, \doi{https://doi.org/10.1016/j.atmosres.2012.05.013}.

\bibitem[{Kumjian et~al.(2021)Kumjian, Lombardo,, and Loeffler}]{Kumjian21}
Kumjian, M.~R., K.~Lombardo, and S.~Loeffler, 2021: {The evolution of hail
  production in simulated supercell storms}. \textit{J. Atmos. Sci.},
  \textbf{78~(11)}, 3417--3440, \doi{10.1175/JAS-D-21-0034.1}.

\bibitem[{Kunz and Kugel(2015)Kunz, and Kugel}]{Kunz15}
Kunz, M., and P.~I.~S. Kugel, 2015: {Detection of hail signatures from
  single-polarization C-band radar reflectivity}. \textit{Atmos. Res.},
  \textbf{153}, 565--577, \doi{10.1016/j.atmosres.2014.09.010}.

\bibitem[{Lakshmanan et~al.(2006)Lakshmanan, Smith, Hondl, Stumpf,, and
  Witt}]{Lakshmanan06}
Lakshmanan, V., T.~Smith, K.~Hondl, G.~J. Stumpf, and A.~Witt, 2006: {A
  real-time, three-dimensional, rapidly updating, heterogeneous radar merger
  technique for reflectivity, velocity, and derived products}. \textit{Wea.
  Forecasting}, \textbf{21~(5)}, 802--823, \doi{10.1175/WAF942.1}.

\bibitem[{Langston et~al.(2007)Langston, Zhang,, and Howard}]{Langston07}
Langston, C., J.~Zhang, and K.~Howard, 2007: {Four-dimensional dynamic radar
  mosaic}. \textit{J. Atmos. Ocean. Technol.}, \textbf{24~(5)}, 776--790,
  \doi{10.1175/JTECH2001.1}.

\bibitem[{Laroche and Zawadzki(1994)Laroche, and Zawadzki}]{Laroche94}
Laroche, S., and I.~Zawadzki, 1994: {A variational analysis method for
  retrieval of three-dimensional wind field from single-Doppler radar data}.
  \textit{J. Atmos. Sci.}, \textbf{51~(18)}, 2664--2682,
  \doi{10.1175/1520-0469(1994)051<2664:AVAMFR>2.0.CO;2}.

\bibitem[{Levizzani and Setv{\'{a}}k(1996)Levizzani, and
  Setv{\'{a}}k}]{Levizzani96}
Levizzani, V., and M.~Setv{\'{a}}k, 1996: {Multispectral, high-resolution
  satellite observations of plumes on top of convective storms}. \textit{J.
  Atmos. Sci.}, \textbf{53~(3)}, 361--369,
  \doi{10.1175/1520-0469(1996)053<0361:MHRSOO>2.0.CO;2}.

\bibitem[{Louf and Protat(2023)Louf, and Protat}]{Louf23}
Louf, V., and A.~Protat, 2023: {Real-time monitoring of weather radar network
  calibration and antenna pointing}. \textit{J. Atmos. Ocean. Technol.},
  \doi{https://doi.org/10.1175/JTECH-D-22-0118.1}.

\bibitem[{Lukach et~al.(2017)Lukach, Foresti, Giot,, and Delobbe}]{Lukach17}
Lukach, M., L.~Foresti, O.~Giot, and L.~Delobbe, 2017: {Estimating the
  occurrence and severity of hail based on 10 years of observations from
  weather radar in Belgium}. \textit{Meteor. Appl.}, \textbf{24~(2)}, 250--259,
  \doi{10.1002/met.1623}.

\bibitem[{May et~al.(2002)May, Jameson, Keenan, Johnston,, and Lucas}]{May02}
May, P.~T., A.~R. Jameson, T.~D. Keenan, P.~E. Johnston, and C.~Lucas, 2002:
  {Combined wind profiler/Polarimetric radar studies of the vertical motion and
  microphysical characteristics of tropical sea-breeze thunderstorms}.
  \textit{Mon. Wea. Rev.}, \textbf{130~(9)}, 2228--2239,
  \doi{10.1175/1520-0493(2002)130<2228:CWPPRS>2.0.CO;2}.

\bibitem[{McAneney et~al.(2019)McAneney, Sandercock, Crompton, Mortlock,
  Musulin, Pielke,, and Gissing}]{Mcaneney19}
McAneney, J., B.~Sandercock, R.~Crompton, T.~Mortlock, R.~Musulin, R.~Pielke,
  and A.~Gissing, 2019: {Normalised insurance losses from Australian natural
  disasters: 1966–2017}. \textit{Environ. Hazards}, \textbf{18~(5)},
  414--433, \doi{10.1080/17477891.2019.1609406}.

\bibitem[{McMaster(2001)}]{Mcmaster01}
McMaster, H., 2001: {Hailstorm risk assessment in rural new south wales}.
  \textit{Nat Hazards}, \textbf{24~(2)}, 187--196,
  \doi{10.1023/A:1011820206279}.

\bibitem[{Mroz et~al.(2017)Mroz, Battaglia, Lang, Cecil, Tanelli,, and
  Tridon}]{Mroz17}
Mroz, K., A.~Battaglia, T.~J. Lang, D.~J. Cecil, S.~Tanelli, and F.~Tridon,
  2017: {Hail-detection algorithm for the GPM core observatory satellite
  sensors}. \textit{J. Appl. Meteor. Climatol.}, \textbf{56~(7)}, 1939--1957,
  \doi{10.1175/JAMC-D-16-0368.1}.

\bibitem[{Murillo and Homeyer(2019)Murillo, and Homeyer}]{Murillo19}
Murillo, E.~M., and C.~R. Homeyer, 2019: {Severe hail fall and hailstorm
  detection using remote sensing observations}. \textit{J. Appl. Meteor.
  Climatol.}, \textbf{58~(5)}, 947--970, \doi{10.1175/JAMC-D-18-0247.1}.

\bibitem[{Murillo et~al.(2021)Murillo, Homeyer,, and Allen}]{Murillo21}
Murillo, E.~M., C.~R. Homeyer, and J.~T. Allen, 2021: {A 23-year severe hail
  climatology using GridRad MESH observations}. \textit{Mon. Wea. Rev.},
  \textbf{149~(4)}, 945--958, \doi{https://doi.org/10.1175/MWR-D-20-0178.1}.

\bibitem[{Ni et~al.(2017)Ni, Liu, Cecil,, and Zhang}]{Ni17}
Ni, X., C.~Liu, D.~J. Cecil, and Q.~Zhang, 2017: {On the detection of hail
  using satellite passive microwave radiometers and precipitation radar}.
  \textit{J. Appl. Meteor. Climatol.}, \textbf{56~(10)}, 2693--2709,
  \doi{10.1175/JAMC-D-17-0065.1}.

\bibitem[{Nisi et~al.(2018)Nisi, Hering, Germann,, and Martius}]{Nisi18}
Nisi, L., A.~Hering, U.~Germann, and O.~Martius, 2018: {A 15-year hail streak
  climatology for the alpine region}. \textit{Q. J. R. Meteor. Soc.},
  \textbf{144~(714)}, 1429--1449, \doi{10.1002/qj.3286}.

\bibitem[{Ortega(2018)}]{Ortega18}
Ortega, K., 2018: {Evaluating multi-radar, multi-sensor products for surface
  hailfall diagnosis}. \textit{Electron. J. Sev. Storms Meteor.},
  \textbf{13~(1)}.

\bibitem[{Ortega et~al.(2006)Ortega, Smith,, and Stumpf}]{Ortega06}
Ortega, K.~L., T.~M. Smith, and G.~J. Stumpf, 2006: {Verification of
  multi-sensor, multi-radar hail diagnosis techniques}. \textit{Symp. on the
  Challenges of Severe convective Storms}.

\bibitem[{Potts et~al.(2000)Potts, Keenan,, and May}]{Potts00}
Potts, R.~J., T.~D. Keenan, and P.~T. May, 2000: {Radar characteristics of
  storms in the Sydney area}. \textit{Mon. Wea. Rev.}, \textbf{128~(9)},
  3308--3319, \doi{10.1175/1520-0493(2000)128<3308:RCOSIT>2.0.CO;2}.

\bibitem[{Prein and Holland(2018)Prein, and Holland}]{Prein18}
Prein, A.~F., and G.~J. Holland, 2018: {Global estimates of damaging hail
  hazard}. \textit{Wea. Clim. Extremes}, \textbf{22}, 10--23,
  \doi{https://doi.org/10.1016/j.wace.2018.10.004}.

\bibitem[{Protat et~al.(2011)Protat, Bouniol, O’Connor, Klein~Baltink,
  Verlinde,, and Widener}]{Protat11}
Protat, A., D.~Bouniol, E.~J. O’Connor, H.~Klein~Baltink, J.~Verlinde, and
  K.~Widener, 2011: {Cloudsat as a global radar calibrator}. \textit{J. Atmos.
  Ocean. Technol.}, \textbf{28~(3)}, 445--452,
  \doi{https://doi.org/10.1175/2010JTECHA1443.1}.

\bibitem[{Protat et~al.(2022)Protat, Louf, Soderholm, Brook,, and
  Ponsonby}]{Protat22}
Protat, A., V.~Louf, J.~Soderholm, J.~Brook, and W.~Ponsonby, 2022: {Three-way
  calibration checks using ground-based, ship-based, and spaceborne radars}.
  \textit{Atmos. Meas. Tech.}, \textbf{15~(4)}, 915--926,
  \doi{10.5194/amt-15-915-2022}.

\bibitem[{Pulkkinen et~al.(2019)Pulkkinen, Nerini, P{\'{e}}rez~Hortal,
  Velasco-Forero, Seed, Germann,, and Foresti}]{Pulkkinen19}
Pulkkinen, S., D.~Nerini, A.~A. P{\'{e}}rez~Hortal, C.~Velasco-Forero, A.~Seed,
  U.~Germann, and L.~Foresti, 2019: {Pysteps: an open-source python library for
  probabilistic precipitation nowcasting (v1.0)}. \textit{Geosci. Model Dev.},
  \textbf{12~(10)}, 4185--4219, \doi{10.5194/gmd-12-4185-2019}.

\bibitem[{Punge et~al.(2018)Punge, Bedka,, and Kunz}]{Punge18}
Punge, H.~J., K.~M. Bedka, and M.~Kunz, 2018: {Continental scale hail frequency
  estimation from geostationary satellite detection}. \textit{EGU General
  Assembly Conference Abstracts}, 15737.

\bibitem[{Punge et~al.(2021)Punge, Bedka, Kunz, Bang,, and Itterly}]{Punge21}
Punge, H.~J., K.~M. Bedka, M.~Kunz, S.~D. Bang, and K.~F. Itterly, 2021:
  {Characteristics of hail hazard in South Africa based on satellite detection
  of convective storms}. \textit{Nat. Hazards Earth Syst. Sci.}, \textbf{2021},
  1--32, \doi{10.5194/nhess-2021-342}.

\bibitem[{Raupach et~al.(2023)Raupach, Soderholm, Protat,, and
  Sherwood}]{Raupach23}
Raupach, T.~H., J.~Soderholm, A.~Protat, and S.~C. Sherwood, 2023: {An improved
  instability–shear hail proxy for Australia}. \textit{Mon. Wea. Rev.},
  \textbf{151~(2)}, 545--567, \doi{https://doi.org/10.1175/MWR-D-22-0127.1}.

\bibitem[{Ravasi and Vasconcelos(2020)Ravasi, and Vasconcelos}]{Ravasi20}
Ravasi, M., and I.~Vasconcelos, 2020: {Pylops—A linear-operator python
  library for scalable algebra and optimization}. \textit{SoftwareX},
  \textbf{11}, 100\,361, \doi{10.1016/J.SOFTX.2019.100361}.

\bibitem[{Riemann-Campe et~al.(2009)Riemann-Campe, Fraedrich,, and
  Lunkeit}]{Riemann09}
Riemann-Campe, K., K.~Fraedrich, and F.~Lunkeit, 2009: {Global climatology of
  convective available potential energy (CAPE) and convective inhibition (CIN)
  in ERA-40 reanalysis}. \textit{Atmos. Res.}, \textbf{93~(1-3)}, 534--545.

\bibitem[{Rudin et~al.(1992)Rudin, Osher,, and Fatemi}]{Rudin92}
Rudin, L.~I., S.~Osher, and E.~Fatemi, 1992: {Nonlinear total variation based
  noise removal algorithms}. \textit{Phys. D: Nonlinear Phenom.},
  \textbf{60~(1-4)}, 259--268.

\bibitem[{Ryzhkov et~al.(2013)Ryzhkov, Kumjian, Ganson,, and
  Khain}]{Ryzhkov13a}
Ryzhkov, A.~V., M.~R. Kumjian, S.~M. Ganson, and A.~P. Khain, 2013:
  {Polarimetric radar characteristics of melting hail. Part I: theoretical
  simulations using spectral microphysical modeling}. \textit{J. Appl. Meteor.
  Climatol.}, \textbf{52~(12)}, 2849--2870, \doi{10.1175/JAMC-D-13-073.1}.

\bibitem[{Saltikoff et~al.(2019)}]{Saltikoff19}
Saltikoff, E., and Coauthors, 2019: {An overview of using weather radar for
  climatological studies: successes, challenges, and potential}. \textit{Bull.
  Amer. Meteor. Soc.}, \textbf{100~(9)}, 1739--1752,
  \doi{10.1175/BAMS-D-18-0166.1}.

\bibitem[{Schmetz et~al.(1997)Schmetz, Tjemkes, Gube,, and
  de~Berg]}]{Schmetz97}
Schmetz, J., S.~A. Tjemkes, M.~Gube, and L.~v. de~Berg], 1997: {Monitoring deep
  convection and convective overshooting with METEOSAT}. \textit{Adv. Space
  Res.}, \textbf{19~(3)}, 433--441,
  \doi{https://doi.org/10.1016/S0273-1177(97)00051-3}.

\bibitem[{Schuster et~al.(2006)Schuster, Blong,, and McAneney}]{Schuster06}
Schuster, S.~S., R.~J. Blong, and K.~J. McAneney, 2006: {Relationship between
  radar-derived hail kinetic energy and damage to insured buildings for severe
  hailstorms in eastern Australia}. \textit{Atmos. Res.}, \textbf{81~(3)},
  215--235, \doi{https://doi.org/10.1016/j.atmosres.2005.12.003}.

\bibitem[{Schuster et~al.(2005)Schuster, Blong,, and Speer}]{Schuster05}
Schuster, S.~S., R.~J. Blong, and M.~S. Speer, 2005: {A hail climatology of the
  greater Sydney area and New South Wales, Australia}. \textit{Int. J.
  Climatol.}, \textbf{25~(12)}, 1633--1650.

\bibitem[{Shapiro et~al.(2009)Shapiro, Potvin,, and Gao}]{Shapiro09}
Shapiro, A., C.~K. Potvin, and J.~Gao, 2009: {Use of a vertical vorticity
  equation in variational dual-Doppler wind analysis}. \textit{J. Atmos. Ocean.
  Technol.}, \textbf{26~(10)}, 2089--2106, \doi{10.1175/2009JTECHA1256.1}.

\bibitem[{Shapiro et~al.(2010)Shapiro, Willingham,, and Potvin}]{Shapiro10b}
Shapiro, A., K.~M. Willingham, and C.~K. Potvin, 2010: {Spatially variable
  Advection correction of radar data. Part II: test results}. \textit{J. Atmos.
  Sci.}, \textbf{67~(11)}, 3457--3470, \doi{10.1175/2010JAS3466.1}.

\bibitem[{Skripnikov{\'{a}} and
  Rez{\'{a}}{\v{c}}ov{\'{a}}(2014)Skripnikov{\'{a}}, and
  Rez{\'{a}}{\v{c}}ov{\'{a}}}]{Skripnikov14}
Skripnikov{\'{a}}, K., and D.~Rez{\'{a}}{\v{c}}ov{\'{a}}, 2014: {Radar-based
  hail detection}. \textit{Atmos. Res.}, \textbf{144}, 175--185.

\bibitem[{Soderholm et~al.(2022)Soderholm, Louf, Brook,, and Protat}]{Level1b}
Soderholm, J., V.~Louf, J.~Brook, and A.~Protat, 2022: {Australian operational
  weather radar level 1b dataset}. National Computing Infrastructure.

\bibitem[{Soderholm et~al.(2017{\natexlab{a}})Soderholm, McGowan, Richter,
  Walsh, Weckwerth,, and Coleman}]{Soderholm17}
Soderholm, J.~S., H.~McGowan, H.~Richter, K.~Walsh, T.~M. Weckwerth, and
  M.~Coleman, 2017{\natexlab{a}}: {An 18-year climatology of hailstorm trends
  and related drivers across southeast Queensland, Australia}. \textit{Q. J. R.
  Meteor. Soc.}, \textbf{143~(703)}, 1123--1135, \doi{10.1002/qj.2995}.

\bibitem[{Soderholm et~al.(2017{\natexlab{b}})Soderholm, Mcgowan, Richter,
  Walsh, Wedd,, and Weckwerth}]{Soderholm17b}
Soderholm, J.~S., H.~A. Mcgowan, H.~Richter, K.~Walsh, T.~Wedd, and T.~M.
  Weckwerth, 2017{\natexlab{b}}: {Diurnal preconditioning of subtropical
  coastal convective storm environments}. \textit{Mon. Wea. Rev.},
  \textbf{145~(9)}, 3839--3859, \doi{10.1175/MWR-D-16-0330.1}.

\bibitem[{Soderholm et~al.(2019)Soderholm, Turner, Brook, Wedd,, and
  Callaghan}]{Soderholm19}
Soderholm, J.~S., K.~I. Turner, J.~P. Brook, T.~Wedd, and J.~Callaghan, 2019:
  {High-impact thunderstorms of the Brisbane metropolitan area}. \textit{J.
  South. Hemisphere Earth Syst. Sci.}, \textbf{69}.

\bibitem[{Str{\v{z}}inar and Skok(2018)Str{\v{z}}inar, and Skok}]{Strzinar18}
Str{\v{z}}inar, G., and G.~Skok, 2018: {Comparison and optimization of
  radar-based hail detection algorithms in Slovenia}. \textit{Atmos. Res.},
  \textbf{203}, 275--285, \doi{https://doi.org/10.1016/j.atmosres.2018.01.005}.

\bibitem[{Stumpf et~al.(2004)Stumpf, Smith,, and Hocker}]{Stumpf04}
Stumpf, G.~J., T.~M. Smith, and J.~Hocker, 2004: {New hail diagnostic
  parameters derived by integrating multiple radars and multiple sensors}.
  \textit{Preprints, 22nd Conf. on Severe Local Storms, Hyannis, MA, Amer.
  Meteor. Soc. P}, Vol.~7.

\bibitem[{Sturman and Tapper(1996)Sturman, and Tapper}]{Sturman96}
Sturman, A.~P., and N.~J. Tapper, 1996: \textit{{The weather and climate of
  Australia and new Zealand}}. Oxford University Press, USA.

\bibitem[{Taszarek et~al.(2020)Taszarek, Allen, P{\'{u}}{\v{c}}ik, Hoogewind,,
  and Brooks}]{Taszarek20}
Taszarek, M., J.~T. Allen, T.~P{\'{u}}{\v{c}}ik, K.~A. Hoogewind, and H.~E.
  Brooks, 2020: {Severe convective storms across Europe and the united states.
  Part II: ERA5 environments associated with lightning, large hail, severe
  wind, and tornadoes}. \textit{J. Clim.}, \textbf{33~(23)}, 10\,263--10\,286,
  \doi{https://doi.org/10.1175/JCLI-D-20-0346.1}.

\bibitem[{Tippett et~al.(2015)Tippett, Allen, Gensini,, and Brooks}]{Tippett15}
Tippett, M.~K., J.~T. Allen, V.~A. Gensini, and H.~E. Brooks, 2015: {Climate
  and hazardous convective weather}. \textit{Curr. Clim. Change Rep.},
  \textbf{1~(2)}, 60--73, \doi{10.1007/s40641-015-0006-6}.

\bibitem[{Troupin et~al.(2012)}]{Troupin12}
Troupin, C., and Coauthors, 2012: {Generation of analysis and consistent error
  fields using the data interpolating variational analysis (DIVA)}.
  \textit{Ocean Model.}, \textbf{52-53}, 90--101,
  \doi{https://doi.org/10.1016/j.ocemod.2012.05.002}.

\bibitem[{Waldvogel et~al.(1978)Waldvogel, Schmid,, and Federer}]{Waldvogel78I}
Waldvogel, A., W.~Schmid, and B.~Federer, 1978: {The kinetic energy of
  hailfalls. Part I: hailstone spectra}. \textit{J. Appl. Meteor.},
  \textbf{17~(4)}, 515--520,
  \doi{10.1175/1520-0450(1978)017<0515:TKEOHP>2.0.CO;2}.

\bibitem[{Walsh et~al.(2016)}]{Walsh16}
Walsh, K., and Coauthors, 2016: {Natural hazards in Australia: storms, wind and
  hail}. \textit{Clim Change}, \textbf{139~(1)}, 55--67,
  \doi{10.1007/s10584-016-1737-7}.

\bibitem[{Wapler and Lane(2012)Wapler, and Lane}]{Wapler12}
Wapler, K., and T.~P. Lane, 2012: {A case of offshore convective initiation by
  interacting land breezes near Darwin, Australia}. \textit{Meteor. Atmos.
  Phys.}, \textbf{115~(3)}, 123--137, \doi{10.1007/s00703-011-0180-6}.

\bibitem[{Warren et~al.(2018)Warren, Protat, Siems, Ramsay, Louf, Manton,, and
  Kane}]{Warren18}
Warren, R.~A., A.~Protat, S.~T. Siems, H.~A. Ramsay, V.~Louf, M.~J. Manton, and
  T.~A. Kane, 2018: {Calibrating ground-based radars against TRMM and GPM}.
  \textit{J. Atmos. Ocean. Technol.}, \textbf{35~(2)}, 323--346,
  \doi{10.1175/JTECH-D-17-0128.1}.

\bibitem[{Warren et~al.(2020)Warren, Ramsay, Siems, Manton, Peter, Protat,, and
  Pillalamarri}]{Warren20}
Warren, R.~A., H.~A. Ramsay, S.~T. Siems, M.~J. Manton, J.~R. Peter, A.~Protat,
  and A.~Pillalamarri, 2020: {Radar-based climatology of damaging hailstorms in
  Brisbane and Sydney, Australia}. \textit{Q. J. R. Meteor. Soc.},
  \doi{10.1002/qj.3693}.

\bibitem[{Wendt and Jirak(2021)Wendt, and Jirak}]{Wendt21}
Wendt, N.~A., and I.~L. Jirak, 2021: {An hourly climatology of operational MRMS
  MESH-diagnosed severe and significant hail with comparisons to storm data
  hail reports}. \textit{Wea. Forecasting}, \textbf{36~(2)}, 645--659,
  \doi{https://doi.org/10.1175/WAF-D-20-0158.1}.

\bibitem[{Wilson et~al.(2009)Wilson, Ortega,, and Lakshmanan}]{Wilson09}
Wilson, C., K.~Ortega, and V.~Lakshmanan, 2009: {Evaluating multi-radar,
  multisensor hail diagnosis with high resolution hail reports}. \textit{25th
  Conf. on Interactive Information Processing Systems}, American Meteorological
  Society, Phoenix, AZ.

\bibitem[{Witt et~al.(1998)Witt, Eilts, Stumpf, Johnson, Mitchell,, and
  Thomas}]{Witt98}
Witt, A., M.~D. Eilts, G.~J. Stumpf, J.~T. Johnson, E.~D.~W. Mitchell, and
  K.~W. Thomas, 1998: {An enhanced hail detection algorithm for the WSR-88D}.
  \textit{Wea. Forecasting}, \textbf{13~(2)}, 286--303,
  \doi{10.1175/1520-0434(1998)013<0286:AEHDAF>2.0.CO;2}.

\bibitem[{York(1966)}]{York66}
York, D., 1966: {Least-squares fitting of a straight line}. \textit{Can. J.
  Phys.}, \textbf{44~(5)}, 1079--1086.

\bibitem[{Zhou et~al.(2021)Zhou, Zhang, Allen, Ni,, and Ng}]{Zhou21}
Zhou, Z., Q.~Zhang, J.~T. Allen, X.~Ni, and C.-P. Ng, 2021: {How many types of
  severe hailstorm environments are there globally?} \textit{Geophys. Res.
  Lett.}, \textbf{48~(23)}, e2021GL095\,485,
  \doi{https://doi.org/10.1029/2021GL095485}.

\end{thebibliography}

\end{document}